\newcommand*{\ATLASLATEXPATH}{latex/}

\documentclass[cernpreprint,texlive=2011,txfonts,UKenglish]{latex/atlasdoc}

\usepackage{\ATLASLATEXPATH atlaspackage}
\usepackage[italic]{hepnames}

\usepackage{\ATLASLATEXPATH atlasbiblatex}

\usepackage{\ATLASLATEXPATH atlasphysics}

\graphicspath{{logos/}{figures/}}

\usepackage{onelepton2015-defs}

\usepackage{multirow}
\usepackage{lscape}
\usepackage{rotating} 

\AtlasTitle{Search for gluinos in events with an isolated lepton, jets and
missing transverse momentum at $\sqrt{s}$ = 13 \TeV\ with the ATLAS detector}

\author{The ATLAS Collaboration}

\PreprintIdNumber{CERN-EP-2016-037}

\AtlasJournal{EPJ C}
\AtlasJournalRef{Eur. Phys. J. C 76 (2016) 565}
\AtlasDOI{10.1140/epjc/s10052-016-4397-x}

\AtlasAbstract{%
The results of a search for gluinos in final states with an isolated electron or
muon, multiple jets and large missing transverse momentum using
proton--proton collision data at a centre-of-mass energy of $\sqrt{s} = 13
\TeV$ are presented.
The dataset used was recorded in 2015 by the ATLAS experiment at the Large
Hadron Collider and corresponds to an integrated luminosity of $3.2$~\ifb. Six signal selections are defined that best exploit the signal characteristics.
The data agree with the Standard Model background expectation in all
six signal selections, and the largest deviation is a $2.1$ standard deviation
excess. The results are interpreted in a simplified model where
pair-produced gluinos decay via the lightest chargino
to the lightest neutralino. In this model,
gluinos are excluded up to masses of approximately $1.6$ \TeV\ depending on the
mass spectrum of the simplified model, thus surpassing the limits of
previous searches.}

\hypersetup{pdftitle={ATLAS document},pdfauthor={The ATLAS Collaboration}}

\begin{document}

\maketitle

\section{Introduction}
\label{sec:intro}
Supersymmetry (SUSY)
\cite{Golfand:1971iw,Volkov:1973ix,Wess:1974tw,Wess:1974jb,Ferrara:1974pu,Salam:1974ig}
is a theoretical framework of physics beyond the Standard Model (SM) that
predicts for each SM particle the existence of a supersymmetric partner differing by half a unit of spin. The partner particles of the SM fermions (quarks and leptons) are
the scalar squarks ($\tilde{q}$) and sleptons~($\tilde{\ell}$).
In the boson sector, the supersymmetric partner of the gluon is the fermionic
gluino ($\tilde{g}$), whereas the supersymmetric partners of the Higgs
(higgsinos) and the electro-weak gauge bosons (winos and bino) mix to form charged and
neutral mass eigenstates called charginos ($\tilde{\chi}^{\pm}_{1,2}$) and
neutralinos ($\tilde{\chi}^{0}_{1,2,3,4}$). In the Minimal Supersymmetric
extension of the Standard Model (MSSM)~\cite{Fayet:1976et,Fayet:1977yc} two
scalar Higgs doublets along with their Higgsino partners are predicted. SUSY
addresses the SM hierarchy
problem~\cite{Sakai:1981gr,Dimopoulos:1981yj,Ibanez:1981yh,Dimopoulos:1981zb} provided that the masses of at least some of the supersymmetric particles (most notably the higgsinos, the top squarks and the gluinos) are near the \TeV\ scale.

In R-parity-conserving SUSY~\cite{Farrar:1978xj}, gluinos might be pair-produced
at the Large Hadron Collider (LHC) via the strong interaction and decay either directly or
via intermediate states to the lightest supersymmetric particle (LSP).
The LSP is stable and is assumed to be only weakly interacting, making it a candidate for
dark matter~\cite{Goldberg:1983nd,Ellis:1983ew}.

This paper considers a SUSY-inspired model where pair-produced gluinos decay
via the lightest chargino~($\tilde{\chi}^{\pm}_{1}$) to the LSP, which is
assumed to be the lightest neutralino (\ninoone). The three-body decay of the gluino to the
chargino proceeds via $\tilde{g} \rightarrow
q\bar{q'} \chinoonepm$. The chargino decays to the LSP by emitting an on-
or off-shell $W$ boson, depending on the available phase space. In the MSSM this
decay chain is realised when the gluino decays via a virtual squark that is the partner of the
left-handed SM quark, to the chargino with a dominant wino component. In the MSSM the mass of the chargino is independent of the mass of the gluino.

The experimental signature characterising this search consists of a lepton
(electron or muon), several jets, and missing transverse momentum (${\boldsymbol
p}_{\mathrm{T}}^\mathrm{miss}$ with magnitude \met) from the undetectable
neutralinos and neutrino(s). The analysis is based on two complementary sets of
search channels. The first set requires a low transverse momentum (\pT) lepton
($7/6 < \pt(e/\mu) < 35 \GeV$), and is referred to as the soft-lepton channel,
while the second set requires a high-$\pT$~lepton ($\pt(e/\mu) > 35 \GeV$) and
is referred to as the hard-lepton channel. The two sets target SUSY models with
small and large mass differences between the predicted supersymmetric particles,
respectively. The search uses the ATLAS data collected in proton--proton LHC
collisions in 2015 corresponding to an integrated luminosity of $3.2$ $\ifb$ at
a centre-of-mass energy of $13$ \TeV.

The analysis extends previous
ATLAS searches with similar event selections~\cite{Aad:2015mia} which were
performed with data collected during the first data-taking campaign between 2010
and 2012 (LHC \mbox{Run 1}) at a centre-of-mass energy of up to $8$ \TeV.
The results of all Run 1 ATLAS searches targeting squark and gluino pair-
production are summarised in Ref.~\cite{Aad:2015iea}. The CMS Collaboration has
performed similar searches for gluinos with decays via intermediate supersymmetric 
particles in Run 1~\cite{Chatrchyan:2013fea,Chatrchyan:2014lfa} and Run 2~\cite{CMS-PAS-SUS-15-006}.

This paper is structured as follows. After a brief description of the ATLAS
detector in Section~\ref{sec:detector}, the simulated data samples for the
background and signal processes used in the analysis as well as the dataset and
the trigger strategy are detailed in Section~\ref{sec:samples}. The
reconstructed objects and quantities used in the analysis are described in
Section~\ref{sec:objects} and the event selection is presented in
Section~\ref{sec:selection}. The background estimation and the systematic
uncertainties associated with the expected event yields are discussed in
Sections~\ref{sec:background} and~\ref{sec:systematics}, respectively, while
details of the statistical interpretation are given in Section~\ref{sec:fit}.
Finally, the results of the analysis are presented in Section~\ref{sec:result}
and are followed by a conclusion.


%
\section{ATLAS detector}
\label{sec:detector}
\interfootnotelinepenalty=10000

ATLAS~\cite{PERF-2007-01} is a general-purpose detector with a forward-backward
symmetric design that provides almost full solid angle coverage around the
interaction point.\footnote{ATLAS uses a right-handed coordinate system with its
origin at the nominal interaction point (IP) in the centre of the detector and
the $z$-axis along the beam pipe. The $x$-axis points from the IP to the centre
of the LHC ring, and the $y$-axis points upward. Cylindrical coordinates
$(r,\phi)$ are used in the transverse plane, $\phi$ being the azimuthal angle
around the $z$-axis. The pseudorapidity is defined in terms of the polar angle
$\theta$ as $\eta=-\ln\tan(\theta/2)$ and the rapidity is defined as $y = 0.5
\ln[(E + p_z)/(E - p_z)]$, where $E$ is the energy and $p_z$ the longitudinal
momentum of the object of interest.}
The main components of ATLAS are the inner detector (ID), which is surrounded by
a superconducting solenoid providing a $2$ T axial magnetic field, the calorimeter
system, and the muon spectrometer (MS), which is immersed in a magnetic field
generated by three large superconducting toroidal magnets. The ID provides track
reconstruction within $|\eta| < 2.5$, employing pixel detectors close to the beam pipe, silicon microstrip detectors at intermediate radii, and a straw-tube tracker with particle identification capabilities based on transition radiation at radii up to $1080$ mm. The innermost pixel detector layer, the insertable B-layer~\cite{B-layerRef}, was added during the shutdown between LHC Run 1 and Run 2, at a radius of $33$ mm around a new, narrower, beam pipe. The calorimeters cover $|\eta| < 4.9$, the forward region ($3.2 < |\eta| < 4.9$) being instrumented with a liquid-argon (LAr) calorimeter for both the
electromagnetic and hadronic measurements. In the central region, a
lead/LAr electromagnetic calorimeter covers $|\eta| < 3.2$,
while the hadronic calorimeter uses two different detector technologies, with
scintillator tiles ($|\eta| < 1.7$) or LAr ($1.5 < |\eta| < 3.2$) as the active
medium. The MS consists of three layers of precision tracking chambers providing coverage
over $|\eta| < 2.7$, while dedicated fast chambers allow triggering over $|\eta|
< 2.4$. The ATLAS trigger system (developed from Ref.~\cite{atlastrigger})
consists of a hardware-based first-level trigger and a software based high-level trigger.

\section{Simulated event samples and data samples}
\label{sec:samples}

The signal model considered in this search is a simplified
model~\cite{Alwall:2008ve,Alwall:2008ag,Alves:2011wf} that has been used in
previous similar ATLAS searches~\cite{Aad:2015mia}.
In this model, exclusive pair-production of gluinos is assumed. The gluinos decay
via an intermediate chargino, here the lightest chargino \chinoonepm, into the
lightest supersymmetric particle, the lightest neutralino
\ninoone. The branching ratio of each supersymmetric particle decay considered is assumed
to be 100\%. Other supersymmetric particles not entering the decay chain
described are not considered in this simplified model and their masses are set
to high values. The gluino decay is assumed to proceed only via virtual first- and second-
generation quarks, hence no bottom or top quarks are produced in the simplified
model. The free parameters of the model are the masses of the gluino
($m_{\tilde{g}}$), the chargino ($m_{\chinoonepm}$), and the neutralino
($m_{\tilde{\chi}^{0}_{1}}$).
Two types of scenarios are considered:
in the first type, the mass of the neutralino is fixed to $60 \GeV$, and the
sensitivity is assessed as a function of the gluino mass and a mass-ratio
parameter defined as $x = (m_{\tilde{\chi}_{1}^{\pm}} -
m_{\tilde{\chi}^{0}_{1}})/(m_{\tilde{g}} - m_{\tilde{\chi}^{0}_{1}})$.
In the second type, $m_{\tilde{g}}$ and $m_{\ninoone}$ are free parameters,
while $m_{\chinoonepm}$ is set to $m_{\chinoonepm} = (m_{\tilde{g}} +
m_{\ninoone})/2$.
The decay topology of the simplified model is illustrated in
Fig.~\ref{fig:feynman}.

\begin{figure}[htb]
\centering
\includegraphics[width=0.45\textwidth]{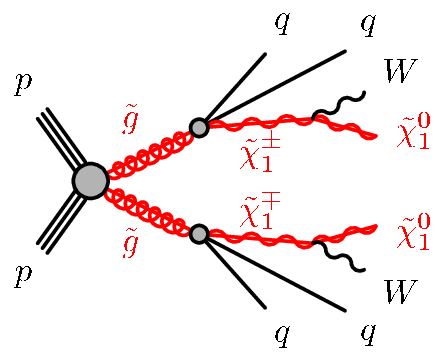}
\caption{The decay topology of the signal model considered in this search.} 
\label{fig:feynman}
\end{figure}

The signal samples are generated using \madgraph 2.2.2~\cite{Alwall:2014hca}
with up to two extra partons in the matrix element, interfaced to
\pythia8.186~\cite{Sjostrand:2007gs} for parton showers and hadronisation. For
the combination of the matrix element and the parton shower the
CKKW-L matching scheme~\cite{CKKW} is applied with a scale parameter that is
set to a quarter of the mass of the gluino. 
The ATLAS A14~\cite{ATL-PHYS-PUB-2014-021} set of tuned parameters (tune) for the underlying event is
used together with the NNPDF2.3~LO~\cite{Ball:2012cx} parton distribution
function (PDF) set. The
EvtGen v1.2.0 program~\cite{EvtGen} is used to describe the properties of the
bottom and charm hadron decays in the signal samples.

The signal cross-sections are calculated at next-to-leading order (NLO) in the
strong coupling constant, adding the resummation of soft gluon emission at
next-to-leading-logarithmic accuracy
(NLL)~\cite{Beenakker:1996ch,Kulesza:2008jb,Kulesza:2009kq,Beenakker:2009ha,Beenakker:2011fu}.
The nominal cross-section and its uncertainty are taken from an envelope of
cross-section predictions using different PDF sets and factorisation and
renormalisation scales~\cite{Kramer:2012bx,Borschensky:2014cia}.

\begin{table*}[ht]
\begin{center}
\footnotesize
\caption{Simulated background event samples:
         the corresponding generator, parton shower, cross-section normalisation,
	 PDF set and underlying-event tune are shown. 
}
\begin{tabular}{l l c c c c }
\hline\hline
Physics process & Generator& Parton shower & Cross-section & PDF set & Tune\\
&& & normalisation & & \\
\hline\hline
$W(\rightarrow \ell\nu)$ + jets              & \sherpa 2.1.1~\cite{Gleisberg:2008ta}    & \sherpa 2.1.1     & NNLO  & NLO CT10    & \sherpa default \\
$Z/\gamma^{*}(\rightarrow \ell \ell)$ + jets & \sherpa 2.1.1   & \sherpa 2.1.1       & NNLO  & NLO CT10         & \sherpa default\\

$t\bar{t}$              & {\sc powheg-box} v2 & \pythia6.428~\cite{Sjostrand:2006za}  & NNLO+NNLL  & NLO CT10  &\sc{Perugia2012}~\cite{Skands:2010ak}\\

Single-top              &&&&\\
($t$-channel)           & {\sc powheg-box} v1 & \pythia6.428 & NLO  & NLO CT10f4   & \sc{Perugia2012}\\
Single-top              &&&&\\
($s$- and $Wt$-channel) & {\sc powheg-box} v2 & \pythia6.428  & NLO  & NLO CT10   & \sc{Perugia2012}\\ 

$t\bar{t}+W/Z/WW$       &  \madgraph 2.2.2 & \pythia8.186   & NLO  & NNPDF2.3LO & A14     \\

$WW$, $WZ$ and $ZZ$     &  \sherpa 2.1.1  & \sherpa 2.1.1      & NLO  & NLO CT10   & \sherpa default \\
\hline\hline
\end{tabular}

\label{tab:MC}
\end{center}
\end{table*}

The simulated event samples for the SM backgrounds are summarised in Table~\ref{tab:MC}, along with the PDFs and tunes used. 
Further samples are also used to assess systematic uncertainties, as explained
in Section~\ref{sec:systematics}.

For the production of \ttbar\ and single top quarks in the $Wt$ and $s$-channel~\cite{ATL-PHYS-PUB-2016-004}
the {\sc powheg-box} v2~\cite{Alioli:2010xd} generator with the CT10~\cite{Lai:2010vv}
PDF sets in the matrix-element calculations is used. Electroweak $t$-channel
single-top-quark events are generated using the {\sc powheg-box} v1 generator. This
generator uses the four-flavour scheme for the NLO matrix-element calculations
together with the fixed four-flavour PDF set CT10f4. For all top-quark processes,
top-quark spin correlations are preserved (for the single-top $t$-channel,
top quarks are decayed using MadSpin~\cite{Artoisenet:2012st}). The parton
shower, fragmentation, and the underlying event are simulated using
\pythia6.428 with the CTEQ6L1~\cite{Pumplin:2002vw} PDF
set and the corresponding {\sc Perugia2012} tune (P2012)~\cite{Skands:2010ak}. The
top-quark mass is assumed to be $172.5 \GeV$. The \ttbar\ events are normalised to the NNLO+NNLL cross-sections.
The single-top-quark events are normalised to the NLO cross-sections.
                         
Events containing $W$ or $Z$ bosons with associated jets ($W$/$Z$+jets)~\cite{ATL-PHYS-PUB-2016-003} are simulated using the
\sherpa2.1.1 generator with massive $b/c$-quarks.
Matrix elements are calculated for up to two partons
at NLO and four partons at leading order (LO). The matrix elements are calculated using the Comix~\cite{Gleisberg:2008fv} and
OpenLoops~\cite{Cascioli:2011va} generators and merged with
the \sherpa2.1.1 parton shower~\cite{Schumann:2007mg} using the ME+PS@NLO
prescription~\cite{Hoeche:2012yf}. The CT10 PDF set is used in conjunction with
a dedicated parton-shower tuning developed by the \sherpa authors. The $W/Z$+jets events are normalised to their NNLO cross-sections~\cite{ATL-COM-PHYS-2010-695}.

The diboson samples~\cite{ATL-PHYS-PUB-2016-002} are generated with the same generator and PDF setup as the
$W/Z$+jets samples described above. The diboson processes
are simulated including final states with four charged leptons, three charged
leptons and one neutrino, two charged leptons and two neutrinos, and one charged lepton and
three neutrinos. The matrix elements contain all diagrams with four
electroweak vertices. They are calculated for up to one parton ($4 \ell$, $2 \ell$+$ 2\nu$) or
no additional partons ($3 \ell$+$1 \nu$, $1 \ell$+$3 \nu$) at NLO and up to three partons
at LO. The diboson cross-sections are taken from the NLO generator used. 

For the $t\bar{t}+W/Z/WW$ processes~\cite{ATL-PHYS-PUB-2016-005}, all events are simulated using \madgraph
2.2.2 at LO interfaced to the \pythia8.186 parton shower
model, with up to two ($t\bar{t}+W$), one ($t\bar{t}+Z$) or no ($t\bar{t}+WW$)
extra partons included in the matrix element. The ATLAS
underlying-event tune A14 is used together with the
NNPDF2.3~LO PDF set. The events are normalised to their
respective NLO cross-sections~\cite{Lazopoulos:2008,Campbell:2012}.

The response of the detector to particles is modelled with a full ATLAS detector
simulation \cite{:2010wqa} using \textsc{Geant4}~\cite{Agostinelli:2002hh}, or
using a fast simulation~\cite{atlfast} based on a parameterisation of the
performance of the electromagnetic and hadronic calorimeters and on \textsc{Geant4}
elsewhere. All background (signal) samples are prepared using the full (fast) detector simulation.

All simulated samples are generated with a varying number of minimum-bias
interactions (simulated using \pythia8 with the MSTW2008LO PDF
set~\cite{Sherstnev:2007nd} and the A2 tune \cite{ATLAS:2012uec}) overlaid on
the hard-scattering event to model the multiple proton--proton interactions in
the same and the nearby bunch crossings.
Corrections are applied to the simulated samples to account for
differences between data and simulation for
trigger, identification and reconstruction efficiencies.

\label{sec:data_trigger}
The proton--proton data analysed in this paper were collected by ATLAS in 2015 at a centre-of-mass energy of $13$~\TeV. During this period the
instantaneous luminosity of the LHC reached $5.0 \times 10^{33}$
cm$^{-2}\rm{s}^{-1}$ with a mean number of additional $pp$ interactions per
bunch crossing of approximately $14$. After application of data-quality
requirements related to the beam and detector conditions, the total integrated
luminosity amounts to $3.2$~\ifb, with an associated uncertainty of $\pm5$\%. 
These values are 
derived following the same methodology as the one detailed in
Ref.~\cite{2015lumi}.

The data are collected using an \met\ trigger with a
threshold of $70 \GeV$. This trigger is close to fully efficient 
after applying the requirement on the offline \met\ to be larger than $200~\GeV$.
%
\section{Object reconstruction and identification}
\label{sec:objects}
The reconstructed primary vertex of an event is required to be consistent with
the interaction region and to have at least two associated tracks with $\pt > 400 \MeV$. When more than one such vertex is found, the
vertex with the largest $\sum p_\mathrm{T}^2$ of the associated tracks is chosen.

In the analysis, a distinction is made between \emph{preselected}
reconstructed objects, which fulfil a set of basic criteria and are used
in the $\met$ computation, and \emph{signal} objects that enter
the various control, validation and signal regions and are subject to more stringent requirements.

Jets are reconstructed from topological clusters in the calorimeters
using the anti-$k_t$ algorithm with a radius parameter
$R=0.4$~\cite{Cacciari:2008gp,ATL-PHYS-PUB-2015-036}.
Prior to jet reconstruction, clusters are calibrated to the electromagnetic scale response. Additional correction factors derived from simulation and data are applied to the measured jet energy to calibrate it to the particle level~\cite{ATL-PHYS-PUB-2015-036}. To mitigate the contributions from pile-up, the median energy density of all the jets in the event, multiplied by the jet area, is subtracted from the reconstructed jet energy~\cite{Cacciari2008119,Aad:2015ina}. Preselected jets are required to have $\pt > 20 \GeV$ and $|\eta| < 4.5$. The contamination from cosmic rays, other sources of non-collision background and detector noise is suppressed using dedicated jet-quality criteria~\cite{ATLAS-CONF-2015-029}: when such criteria are not fulfilled, the event is rejected. 

Electron candidates are reconstructed using ID tracks matched to energy clusters
in the electromagnetic calorimeter. They are identified according to the
likelihood-based \emph{loose} criteria~\cite{ATL-PHYS-PUB-2015-041}. Preselected electrons in the soft-lepton (hard-lepton) channel must satisfy $p_\text{T}>7 (10) \GeV$ and $|\eta|<2.47$.
When the angular separation
$\Delta R=\sqrt{(\Delta y)^2+(\Delta\phi)^2}$ between an electron candidate
and a preselected jet amounts to $0.2<\Delta R(e,\mathrm{jet})<0.4$, the jet is
retained and the electron is rejected to remove electrons originating from $b$-hadron decays. Since all electrons are also
reconstructed as jets, if $\Delta R(e,\mathrm{jet})<0.2$ the electron is kept
and the jet is discarded. Finally, electron candidates with a $\Delta
R(e,\mu)<0.01$ with respect to a preselected muon (defined below) are rejected and the muon is kept
to suppress the contribution of electron candidates from muon bremsstrahlung and subsequent photon conversion.

Muon candidates are reconstructed by combining tracks formed in the ID and the MS
sub-systems. The \emph{Medium} identification criteria are applied, which offer
good efficiency and purity for the selected muons \cite{ATL-PHYS-PUB-2015-037}.
Preselected muons in the soft-lepton (hard-lepton) channel are required to have $\pt>6
(10) \GeV$ and $|\eta|<2.4$. Muons with an angular separation of $\Delta
R(\mu,\mathrm{jet})<0.4$ with respect to the closest preselected jet are
rejected, after the electron--jet overlap ambiguities are resolved. However, if
the number of tracks with $\pt > 500 \MeV$ associated with the jet is less than three the jet is
discarded and the muon kept.

The $\met$ is calculated as the magnitude of the negative vector sum of the
transverse momenta of identified and calibrated muons, electrons, jets and
photons, in addition to the \emph{soft-track} term. The soft-track term is
defined as the vectorial sum of the ${\boldsymbol
p}_{\mathrm{T}}$ of all reconstructed tracks associated
with the primary vertex that are not associated with the identified objects entering
explicitly the $\met$
computation~\cite{ATL-PHYS-PUB-2015-023,ATL-PHYS-PUB-2015-027}.

Signal jets are required to have $\pt>25 \GeV$ and $|\eta|<2.8$. A likelihood 
discriminant, the \emph{jet-vertex tagger} (JVT), is
used to remove the residual contamination of pile-up jets. The JVT is constructed
from track-based variables that are sensitive to the vertex of origin of the
jet~\cite{ATLAS-CONF-2014-018}. Jets with $\pt<50
\GeV$, $|\eta|<2.4$, and JVT score less than $0.64$ are rejected.

Signal jets containing $b$-hadrons are identified using the \emph{MV2c20}
algorithm \cite{ATL-PHYS-PUB-2015-022} and are hereafter referred to as $b$-tagged
jets. The \emph{MV2c20} algorithm uses as input the impact parameters of all
associated tracks and any reconstructed secondary vertex. The requirement chosen
in the analysis provides an inclusive $b$-tagging efficiency of $77$\% in
simulated $t\bar{t}$ events, along with a rejection factor of $140$ for gluon and
light-quark jets and of $4.5$ for charm
jets~\cite{ATL-PHYS-PUB-2015-022,ATL-PHYS-PUB-2015-039}.

Signal muons and electrons in the soft-lepton and hard-lepton channels are
subject to an additional $\pt<35 \GeV$ or $\pt\geq 35 \GeV$ requirement,
respectively. Electrons must satisfy likelihood-based \emph{tight}
criteria which are defined in Ref.~\cite{ATL-PHYS-PUB-2015-041}. In both
channels, signal leptons must satisfy the
\emph{GradientLoose}~\cite{ATL-PHYS-PUB-2015-037} isolation requirements, which
rely on the use of tracking-based and calorimeter-based variables and
implement a set of $\eta$- and $\pt$-dependent criteria. The efficiency for
prompt leptons with transverse momentum $< 40$
\GeV\ to satisfy the \emph{GradientLoose} requirements is measured
to be about $95$\% in $Z$~$\rightarrow$ $\ell\ell$ events, progressively rising up to $99$\% at 100 \GeV~\cite{ATL-PHYS-PUB-2015-037}.

To enforce compatibility with the primary vertex, the distance $|z_0\cdot
\mathrm{sin}(\theta)|$ is required to be less than $0.5~\mathrm{mm}$ for signal lepton tracks, where $z_0$ is the longitudinal impact
parameter with respect to the primary-vertex position. Moreover, in the transverse plane the distance of
closest approach of the lepton track to the proton beam line, divided by the
corresponding uncertainty, must be less than three for muons
and less than five for electrons. 

Reconstruction, identification and isolation efficiencies in simulation,
when applicable, are calibrated to data for
all reconstructed objects.

\section{Event selection}
\label{sec:selection}

Events selected by the trigger are further required to have a reconstructed primary vertex. An event is rejected if it contains a preselected jet which fails to satisfy the quality 
criteria designed to suppress non-collision backgrounds and detector noise \cite{ATLAS-CONF-2015-029}. 
Exactly one signal lepton is required in both the soft- and hard-lepton channels. Any event with additional preselected leptons is vetoed to suppress the dilepton $t\bar{t}$, single-top ($Wt$-channel) and diboson backgrounds.

A dedicated optimisation study was performed to design signal region (SR) selection criteria and to maximise the signal sensitivity.
Four hard-lepton signal regions and two soft-lepton signal regions are defined, targeting different mass hierarchy scenarios in the simplified model. The selection criteria used to define the signal regions are summarised in Table~\ref{tab:softlepSR} for the soft-lepton channel and in Table~\ref{tab:hardlepSR} for the hard-lepton channel.

\begin{table}[b]
  \small
  \begin{center}
    \caption{Overview of the selection criteria for the soft-lepton signal regions. The symbol \ptl\ refers to signal leptons.}
    \begin{tabular}{lcc}
      \hline\hline
      & \hspace*{30mm} \\[-3mm]
      & {\bf 2-jet soft-lepton SR} & {\bf 5-jet soft-lepton SR}  \\[1mm]
      \hline\hline
      $N_{\mathrm{lep}}$(\ptl$^{=e(\mu)} > 7(6)$\GeV )        & $= 1$           & $=  1$     \\

      \ptl$^{=e(\mu)}$ (\GeV)                    & $7(6)$--$35$   & $7(6)$--$35$    \\
      \hline
      $N_{\mathrm{jet}}$                        & $\geq 2$ & $\geq 5$
      \\
      $p_{\mathrm{T}}^{\mathrm{jet}}$ (\GeV)              & $>$
      180, 30 & $>$ 200, 200, 200, 30, 30  \\
      \hline
      \met\ (\GeV)                              & $> 530$ & $> 375$
      \\
      \mt\ (\GeV)                               & $> 100$ & -
      \\
      \met/$m_{\mathrm{eff}}^{\mathrm{inc}}$   &  $> 0.38$ & -
      \\
      $H_{\mathrm{T}}$ (\GeV) & - & $> 1100$
      \\
      Jet aplanarity                                    & - &  $> 0.02$
      \\
      \hline\hline
    \end{tabular}

    \label{tab:softlepSR}
  \end{center}
\end{table}

\begin{table}[tb]
  \small
  \begin{center}
    \caption{Overview of the selection criteria for the hard-lepton signal
    regions. The symbol \ptl\ refers to signal leptons.
    The mass-ratio parameter $x$ used in the signal region labels is
    defined in Section~\ref{sec:samples}.}
    \begin{tabular}{lcccc}
      \hline\hline
      & \hspace*{30mm} \\[-3mm]
      & {\bf 4-jet high-$x$ SR} & {\bf 4-jet low-$x$ SR} & {\bf 5-jet SR} & {\bf 6-jet SR} \\[1mm]
      \hline\hline
      $N_{\mathrm{lep}}$(\ptl$^{=e(\mu)} > 10$\GeV )      & $= 1$     & $= 1$  &  $= 1$        & $= 1$                        \\
      \ptl$^{=e(\mu)}$ (\GeV)                  & $> 35$   & $> 35$                 & $> 35$                  & $> 35$                      \\
      \hline
      $N_{\mathrm{jet}}$                        & $\geq 4$ & $\geq 4$               & $\geq 5$                & $\geq 6$                    \\
     $p_{\mathrm{T}}^{\mathrm{jet}}$ (\GeV)              & $> 325, 30,... , 30$   & $> 325, 150,... , 150$ & $> 225, 50,... , 50$ & $> 125, 30,... , 30$ \\
      \hline
      \met\ (\GeV)                              & $> 200$  & $> 200$                & $> 250$                 & $> 250$                     \\
      \mt\ (\GeV)                               & $> 425$  & $> 125$                & $> 275$                 & $> 225$                     \\
      \met/$m_{\mathrm{eff}}^{\mathrm{inc}}$   & $> 0.3$  & -                      & $> 0.1$                 & $> 0.2$                     \\
      $m_{\mathrm{eff}}^{\mathrm{inc}}$ (\GeV) & $> 1800$ & $> 2000$               & $> 1800$                & $> 1000$                    \\
      Jet aplanarity                            & -        & $> 0.04$               & $> 0.04$                & $> 0.04$                    \\
      \hline\hline      
    \end{tabular}

\label{tab:hardlepSR}
\end{center}
\end{table}

The observables defined below are used in the event selection.

The transverse mass ($m_{\rm{T}}$) of the lepton and the ${\boldsymbol p}_{\mathrm{T}}^\mathrm{miss}$ is defined as
\begin{equation}
    m_{\mathrm{T}} =
\sqrt{2 p_{\mathrm{T}}^{\ell} \met
  (1-\cos[\Delta\phi({\boldsymbol p}_{\mathrm{T}}^{\ell},{\boldsymbol p}_{\mathrm{T}}^\mathrm{miss})])},
\end{equation}
where $\Delta\phi({\boldsymbol p}_{\mathrm{T}}^{\ell},{\boldsymbol p}_{\mathrm{T}}^\mathrm{miss})$ is the azimuthal angle
between the lepton and the missing transverse momentum.
This is used in the soft-lepton 2-jet signal region and all hard-lepton signal regions to reject $W$+jets and semileptonic \ttbar\ events.

The inclusive effective mass ($m_{\rm{eff}}^{\rm{inc}}$) is the scalar sum of
the \pt~of the signal lepton and jets and the \met:
\begin{equation}
    m_{\mathrm{eff}}^{\mathrm{inc}} = 
   p_{\mathrm{T}}^{\ell} +  \sum_{j=1}^{{N}_\mathrm{jet}}
    p_{\mathrm{T},j} + \met,
\end{equation}
where the index $j$ runs over all the signal jets in the event with $\pt > 30 \GeV$. 
The inclusive effective mass provides good discrimination against SM backgrounds, without being too sensitive to the details of the SUSY cascade decay chain.

The transverse momentum scalar sum ($H_{\mathrm{T}}$) is defined as
\begin{equation}
  H_{\mathrm{T}}= p_{\mathrm{T}}^{\ell} +  \sum_{j=1}^{{N}_\mathrm{jet}}p_{\mathrm{T},j},
\end{equation}
where the index $j$ runs over all the signal jets in the event. The $H_{\mathrm{T}}$ variable is used to define the soft-lepton 5-jet signal region, as the many energetic jets in the signal model render this variable useful to separate signal from background.
 
The ratio \met/$m_{\rm{eff}}^{\mathrm{inc}}$ is used in both the soft- and hard-lepton channels; it provides good discrimination power between signal and background with fake \met\ ~due to instrumental effects.

Additional suppression of background processes is based on the aplanarity variable,
 which is defined as $\cal A$ = $\frac {3}{2}$ $\lambda_{3}$,
where $\lambda_{3}$ is the smallest eigenvalue of the normalised momentum tensor of the jets~\cite{aplanarity:2012}.
Typical measured values lie in the range $0 \leqslant {\cal A} < 0.3$, with values near zero indicating relatively planar background-like events.

The hard-lepton 5-jet region targets scenarios with high gluino masses
and low $\tilde{\chi}^{0}_{1}$ masses in models with the chargino mass
$m_{\tilde{\chi}_{1}^{\pm}}$ chosen such that the mass-ratio parameter $x =
1/2$.
Tight requirements on \mt\ and $m_{\rm{eff}}^{\mathrm{inc}}$ are applied. For the same set of models, the hard-lepton 6-jet region is designed to provide sensitivity to scenarios where the mass separation between the gluino and the neutralino is smaller. For this reason, the requirements on \mt\ and $m_{\rm{eff}}^{\mathrm{inc}}$ are relaxed with respect to the hard-lepton 5-jet region.
Two distinct hard-lepton 4-jet regions are used, both designed to target models where the neutralino mass is fixed to $60$ \GeV, while the gluino mass and the mass-ratio $x$ vary. The 4-jet high-$x$ region is designed for regions of the parameter space where the $W$ boson produced in the chargino decay is significantly boosted, leading to high-$\pt$ leptons. The main characteristics of signal events in the phase-space of this model are large \mt\ values and relatively soft jets emitted from the gluino decays. In the 4-jet low-$x$ region, the $W$ boson tends to be virtual while the jets from the gluino decays tend to have high $\pt$ due to the large gluino--chargino mass difference. Therefore, the \mt\ requirement is relaxed and more stringent jet $\pt$ requirements are imposed.

The soft-lepton channels focus on models with compressed mass spectra. The
soft-lepton 2-jet region provides sensitivity to scenarios characterised
by a relatively heavy neutralino and a small mass separation between the gluino,
the chargino, and the neutralino. Due to the small mass separation, most of the
decay products tend to be low $\pt$, or soft. Thus, a high-$\pt$ initial-state radiation (ISR)
jet recoiling against the rest of the event is required, in order to enhance the kinematic properties of the signal and to provide separation with respect to the
backgrounds.
The soft-lepton 5-jet region is designed to be sensitive to the configurations
in parameter space with a large mass gap between the gluino and chargino and a
small separation between $m_{\tilde{\chi}_{1}^{\pm}}$ and $m_{\tilde{\chi}^{0}_{1}}$. As a consequence, several energetic jets from the
decay of the two gluinos to the charginos are expected, while the virtual $W$ bosons produced
in the decay of the charginos result in low-\pT\
jets and leptons.

%
\section{Background estimation}
\label{sec:background}

The two dominant background processes in final states with one isolated lepton,
multiple jets and large missing transverse momentum are \ttbar\ and
$W$+jets.
The differential distributions arising from these two background processes as
predicted from simulation are simultaneously normalised to the number of data events observed in
dedicated control regions (CR), through the fitting procedure 
explained in Section~\ref{sec:fit}. The simulation is then used to 
extrapolate the measured background rates to the corresponding signal region. 

The control regions are designed to have high purity in the process of
interest, a small contamination from the signal model and enough events to result in
a small statistical uncertainty in the background prediction.
Moreover, they are designed to have kinematic properties resembling as closely as possible those of the
signal regions,  in order to provide good estimates of the
kinematics of background processes there. This procedure limits the impact of potentially 
large systematic uncertainties in the expected yields. 

Additional sources of background events are single-top events ($s$-channel,
$t$-channel and associated production with a $W$ boson), $Z$+jets and diboson
processes ($WW$, $WZ$, $ZZ$, $W\gamma$, $Z\gamma$), and $t\bar{t}$ production in
association with a $W$ or a $Z$ boson. Their contributions are estimated
entirely using simulated event samples normalised to the most accurate theoretical
cross-sections available.

The contribution from multi-jet processes with a misidentified lepton is found to be
negligible once lepton isolation criteria and a stringent \met\ requirement are
imposed. A data-driven matrix method, following the implementation described in
Ref.~\cite{Aad:2015mia}, confirms this background is consistent with zero. This
is mainly a result of the improved lepton reconstruction and identification and the higher threshold on \met~with respect to the previous searches
performed in this final state~\cite{Aad:2015mia}. As this background is found to be negligible it is ignored in all aspects of the analysis.

Figure~\ref{fig:hardSR} visualises the criteria that define the control regions
in the soft-lepton and hard-lepton channels.
Based on these, separate control regions are defined to extract the normalisation factors for $t\bar{t}$ and $W$+jets by requiring at least one, or no, $b$-tagged signal jets, respectively. The cross-contamination between these two types of control regions is accounted for in the fit.

\begin{figure}[htbp]
\centering
\vspace{-0.5cm}
\begin{minipage}{0.46\textwidth}
 \includegraphics[width=\textwidth]{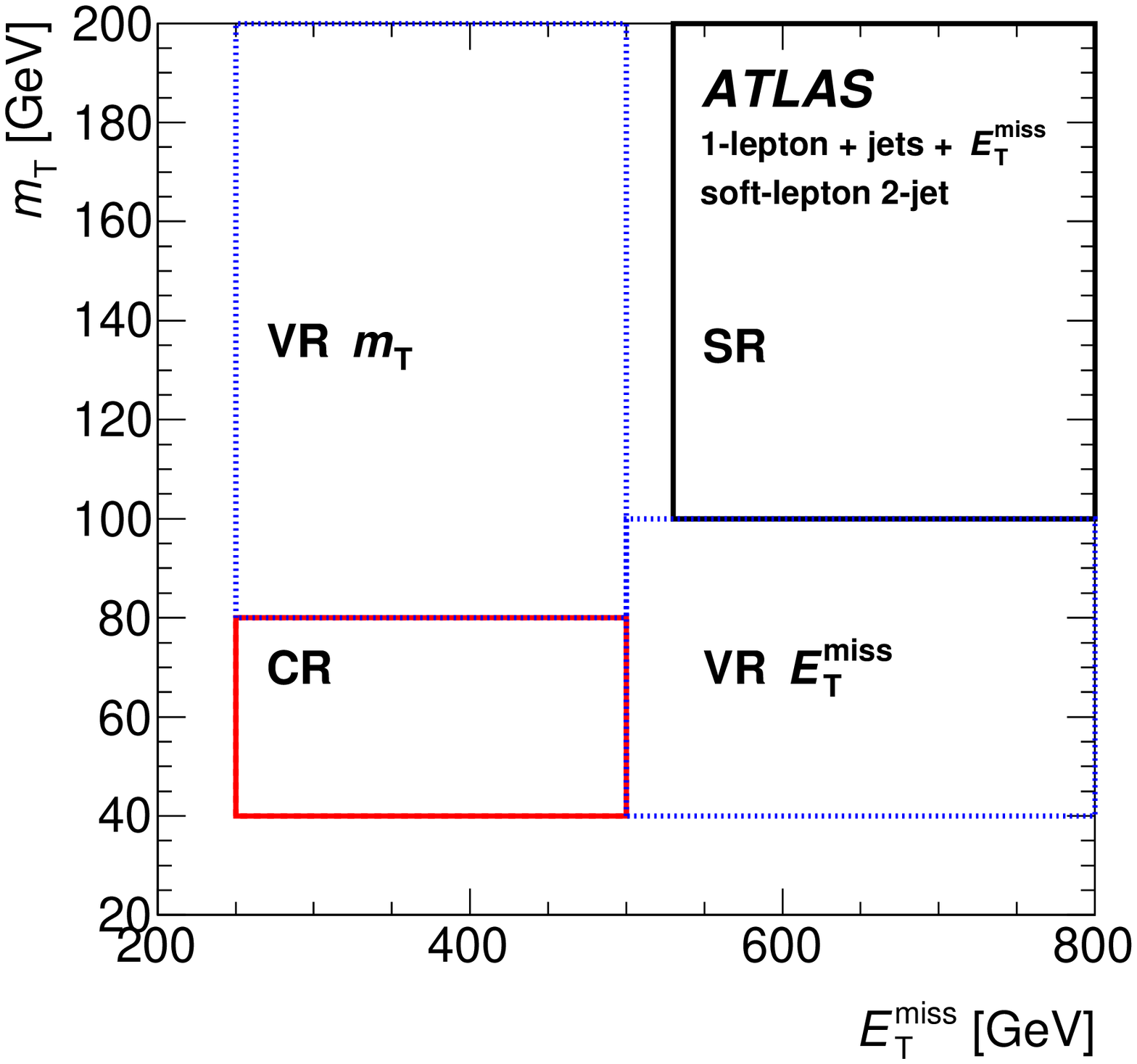}
\end{minipage}
\begin{minipage}{0.46\textwidth}
 \includegraphics[width=\textwidth]{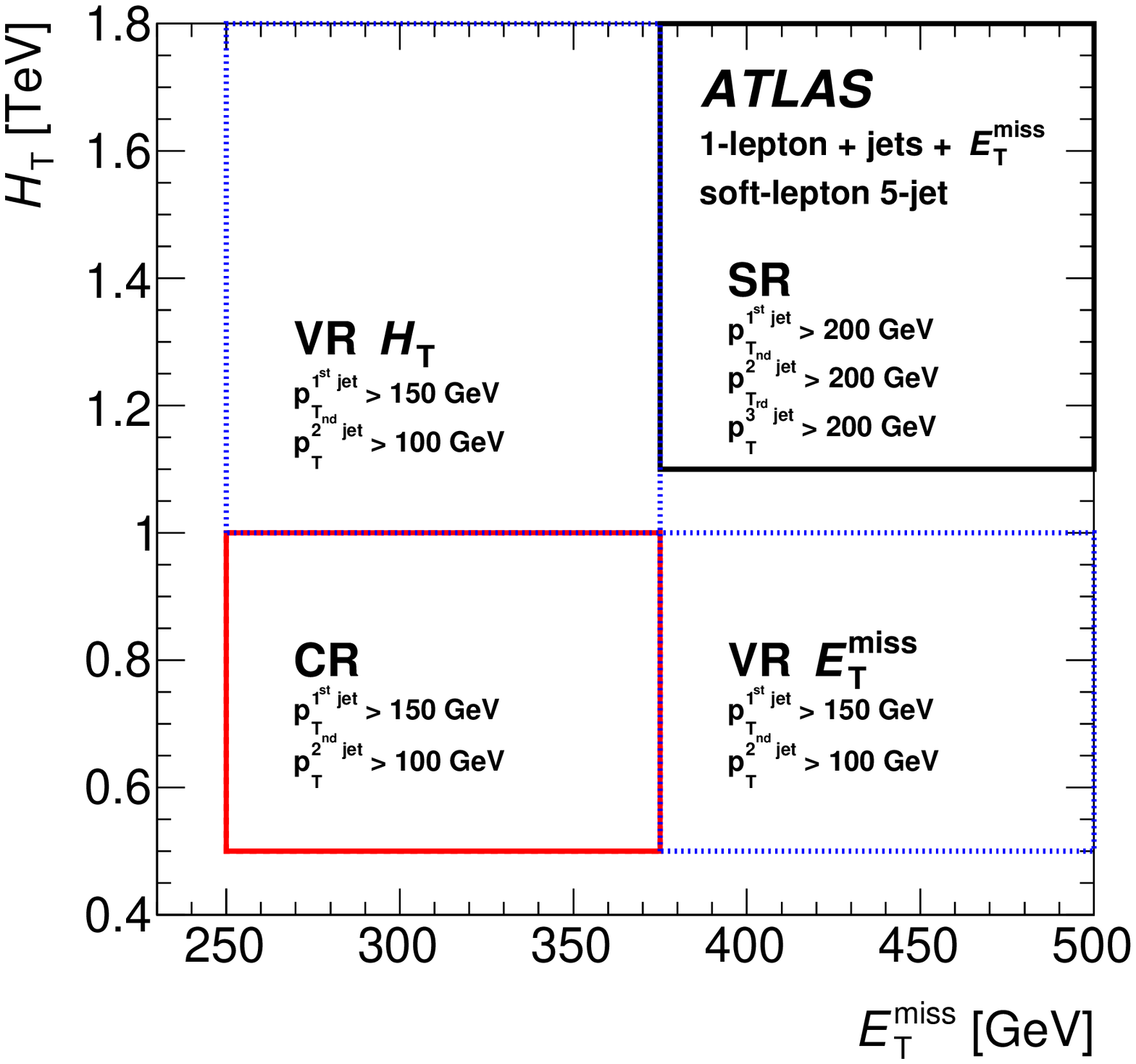}
\end{minipage}\\
\vspace{-0.05cm}
\begin{minipage}{0.46\textwidth}
 \includegraphics[width=\textwidth]{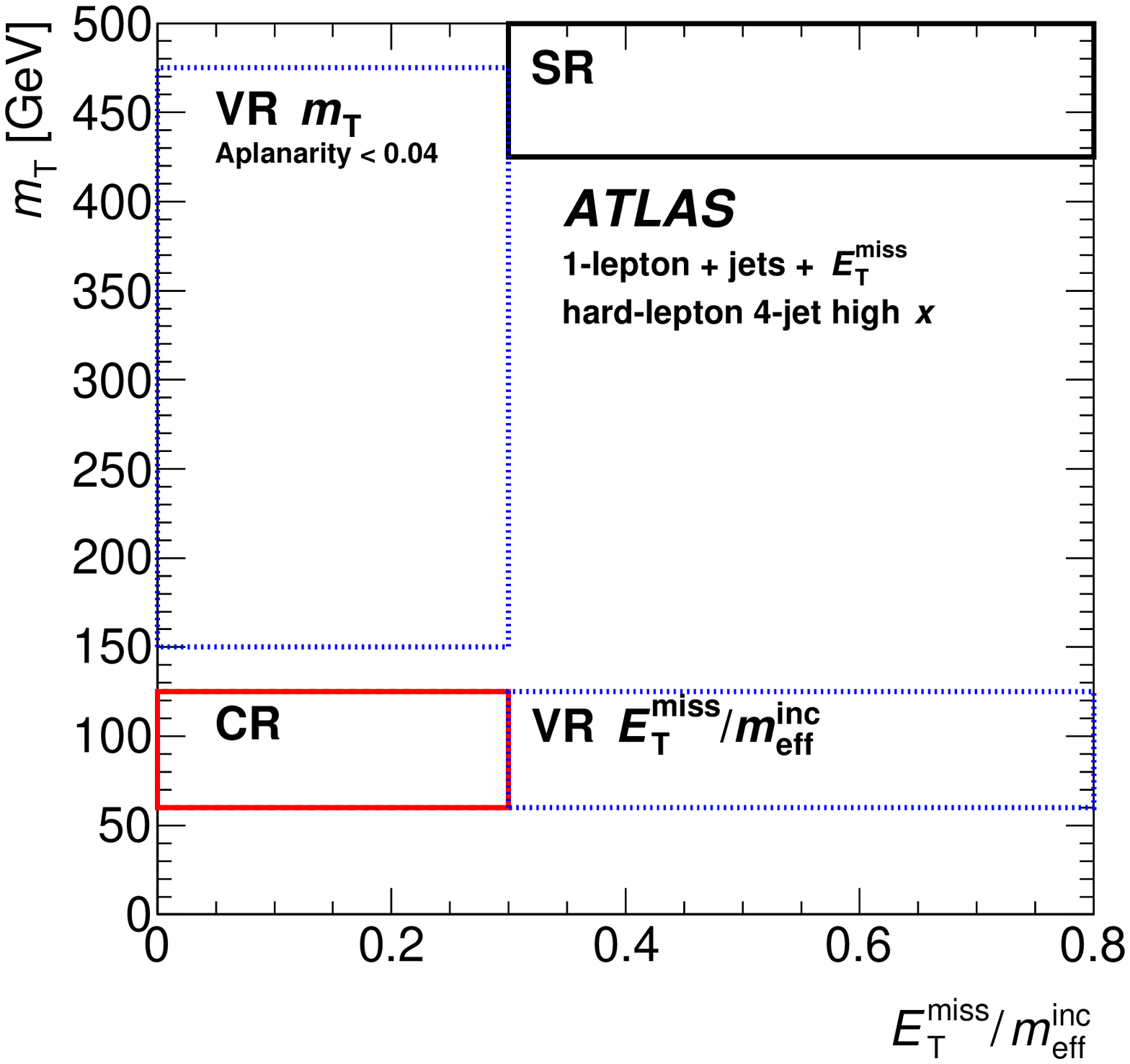}
\end{minipage}
\begin{minipage}{0.46\textwidth}
 \includegraphics[width=\textwidth]{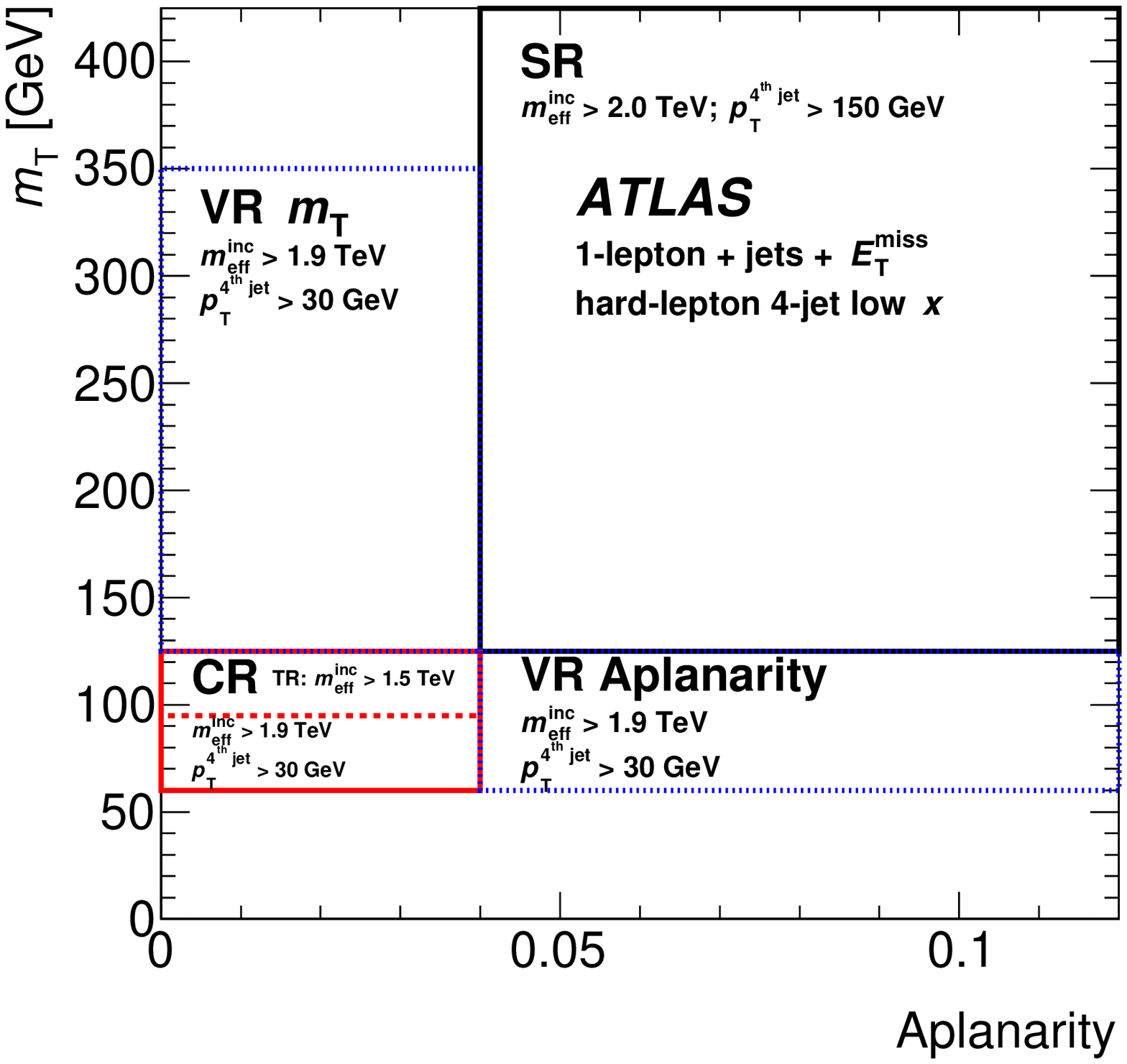}
\end{minipage}\\
\vspace{-0.05cm}
\begin{minipage}{0.46\textwidth}
 \includegraphics[width=\textwidth]{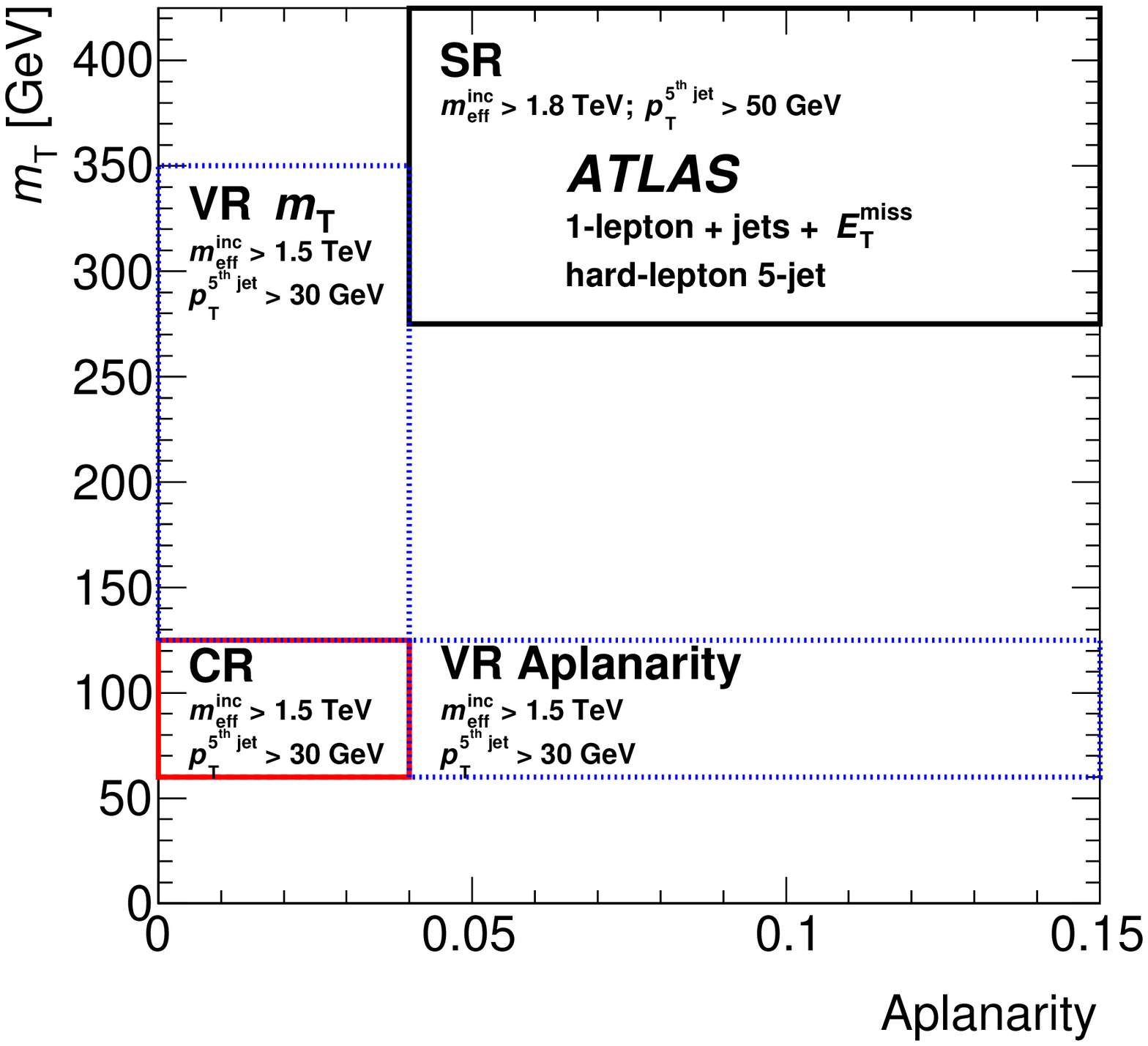}
\end{minipage}
\begin{minipage}{0.46\textwidth}
 \includegraphics[width=\textwidth]{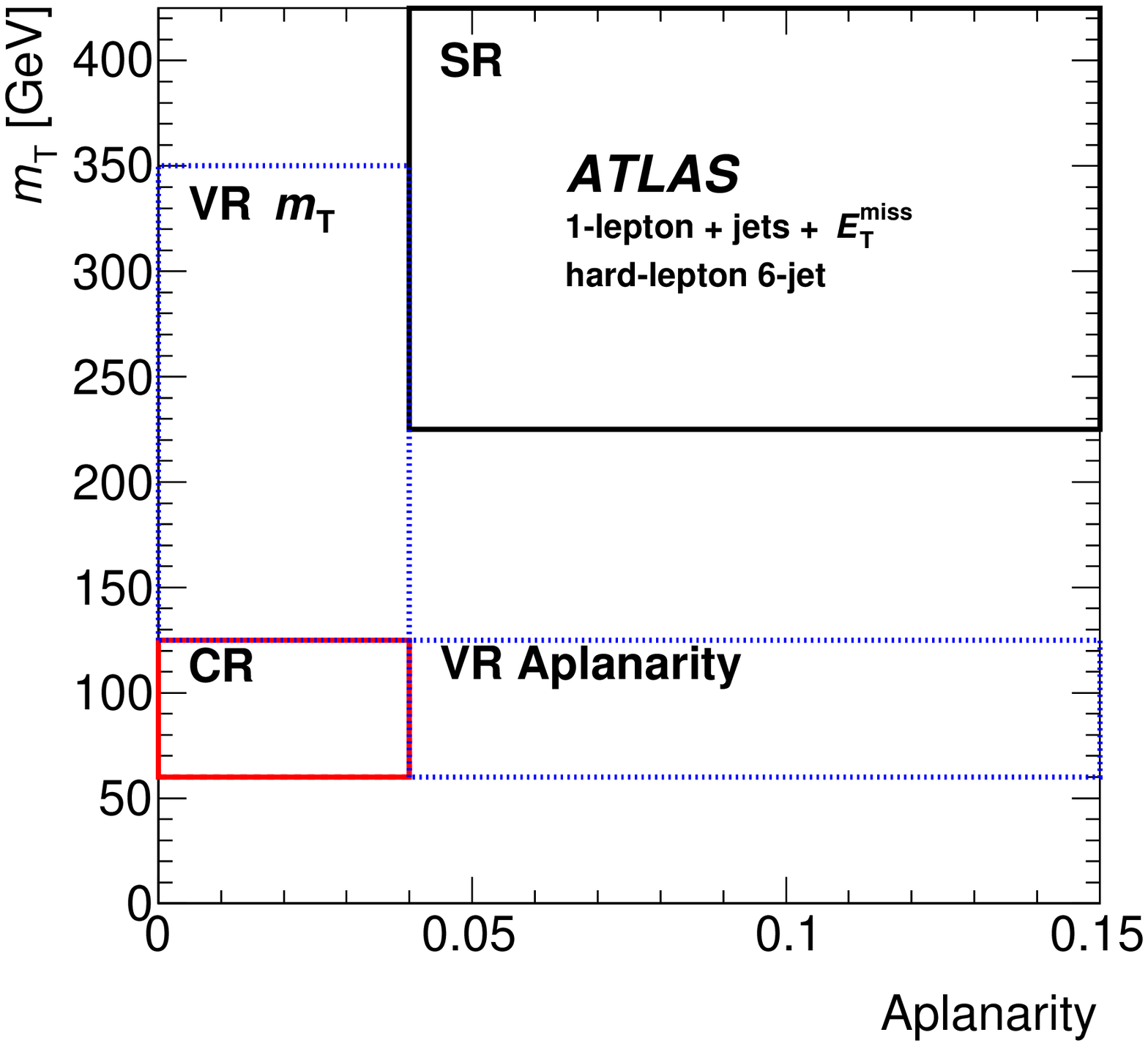}
\end{minipage}
\caption{ Graphical illustration of the soft-lepton 2-jet (top left),
soft-lepton 5-jet (top right), hard-lepton 4-jet high-$x$ (middle left), 4-jet
low-$x$ (middle right), 5-jet (bottom left) and 6-jet (bottom right) signal
(SR), control (CR) and validation (VR) regions. In addition to the two
variables shown on the $x$ and $y$ axes, labels indicate other event selections
that differ between the corresponding control regions, validation regions, and
signal regions. The control regions exist in two
variants:
the \ttbar\ control regions require at least one $b$-tagged jet, while no
$b$-tagged jets are required in the $W$+jets control regions.
} 
\label{fig:hardSR}
\end{figure}

Figure~\ref{sec:background:CRplots} shows the \met\ distribution in selected
soft-lepton and hard-lepton control regions. 
The normalisation of the $W$+jets and \ttbar\ simulations are adjusted to match the observed number of data events in the control region, so that the plots illustrate the modelling of the shape of each variable's distribution. In general, good agreement between data and background simulations is found within the uncertainties in all the control regions used in the analysis.

\begin{figure}
 \begin{center}
  \begin{minipage}{0.49\textwidth}
   \includegraphics[width=\textwidth]{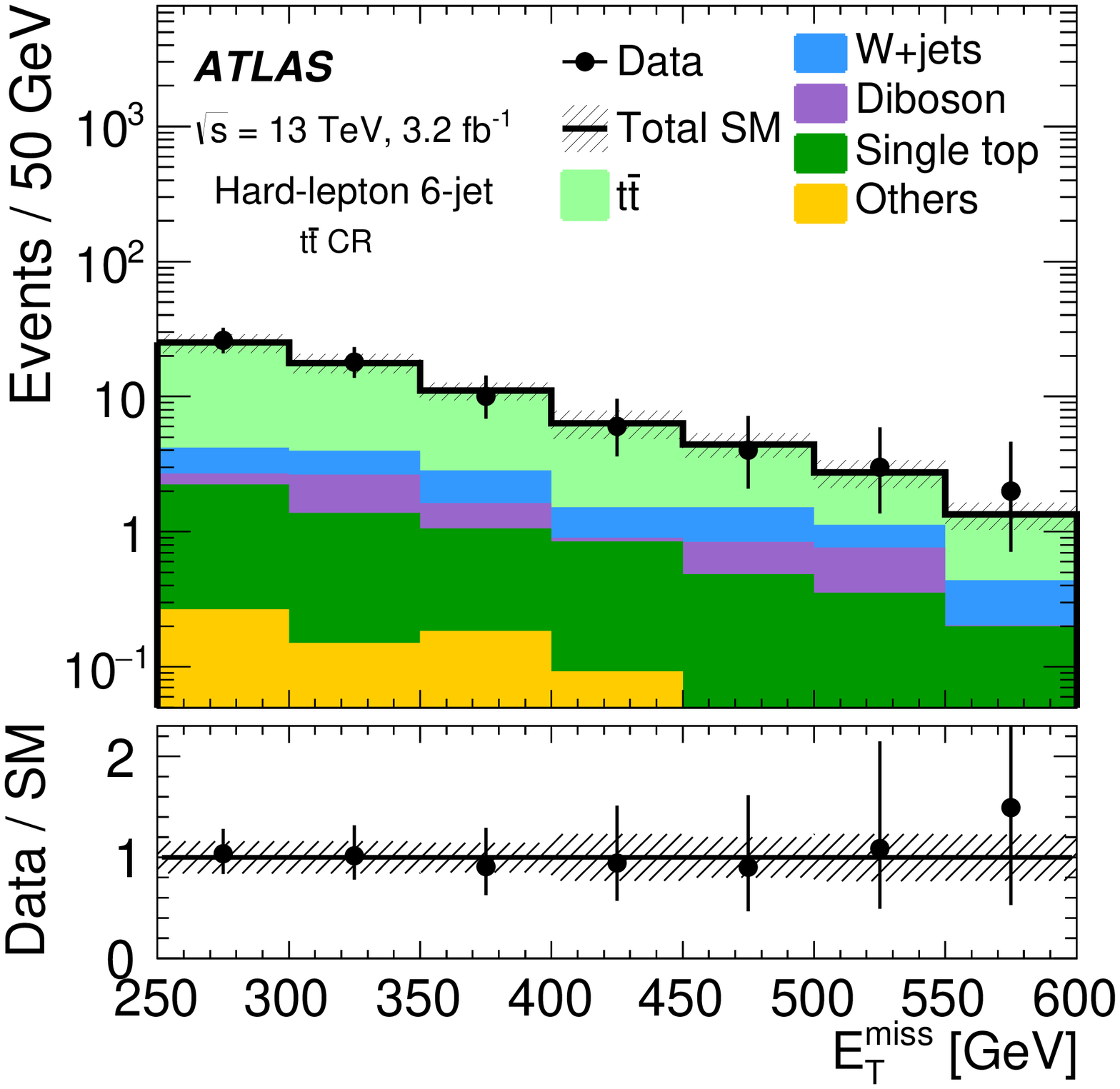}
  \end{minipage}
  \begin{minipage}{0.49\textwidth}
   \includegraphics[width=\textwidth]{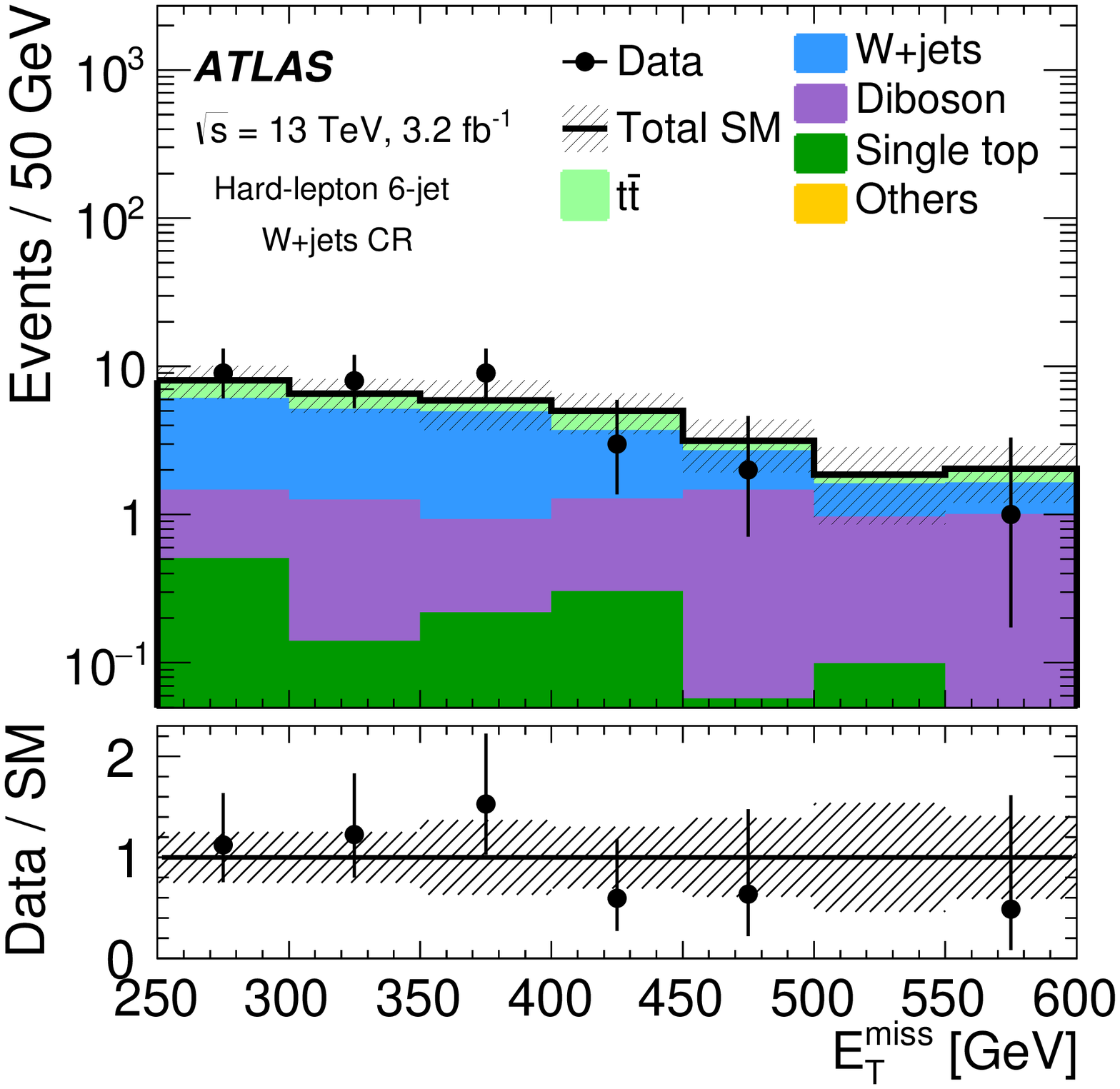}
  \end{minipage}
  \begin{minipage}{0.49\textwidth}
   \includegraphics[width=\textwidth]{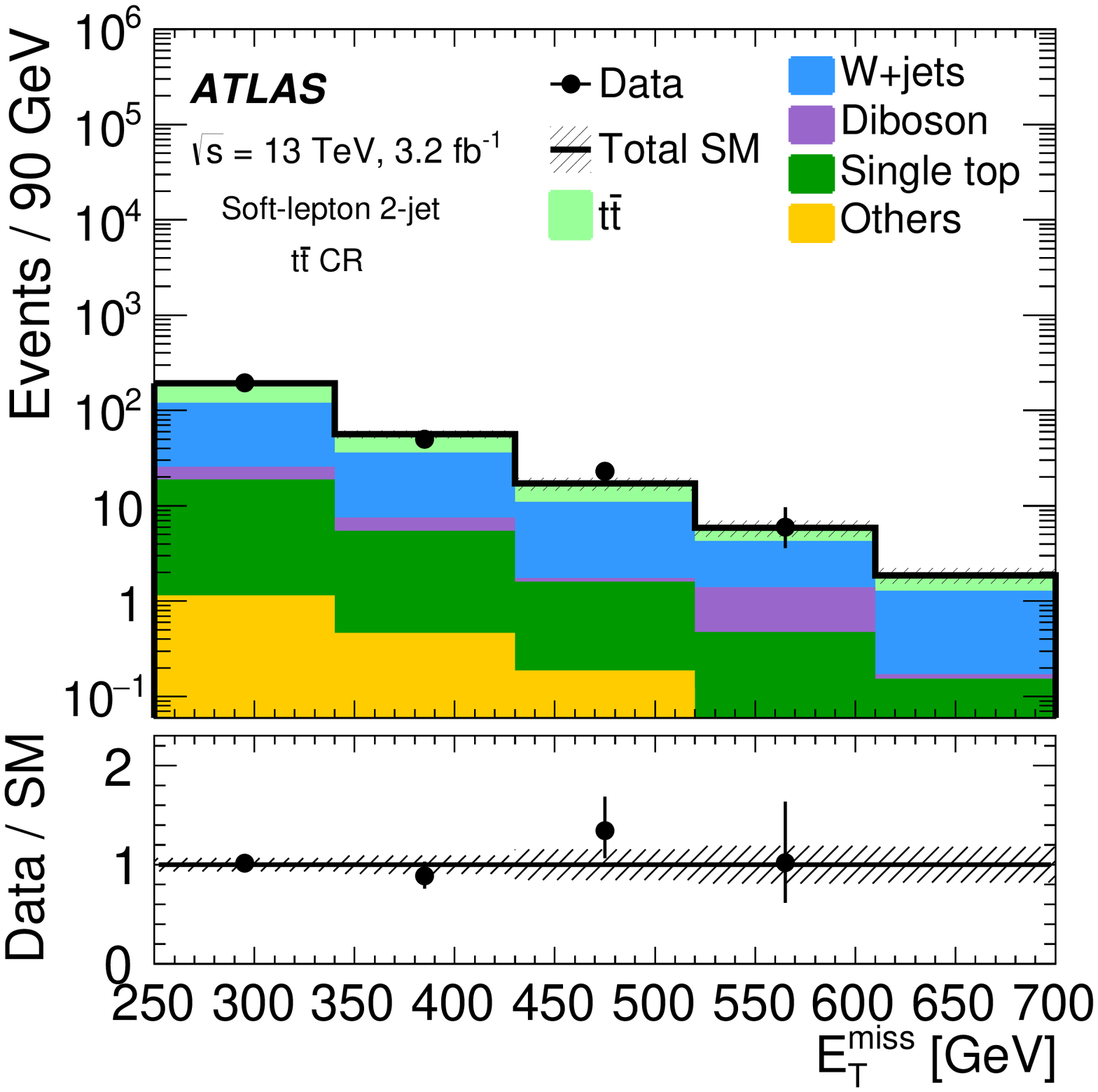}
  \end{minipage}
  \begin{minipage}{0.49\textwidth}
   \includegraphics[width=\textwidth]{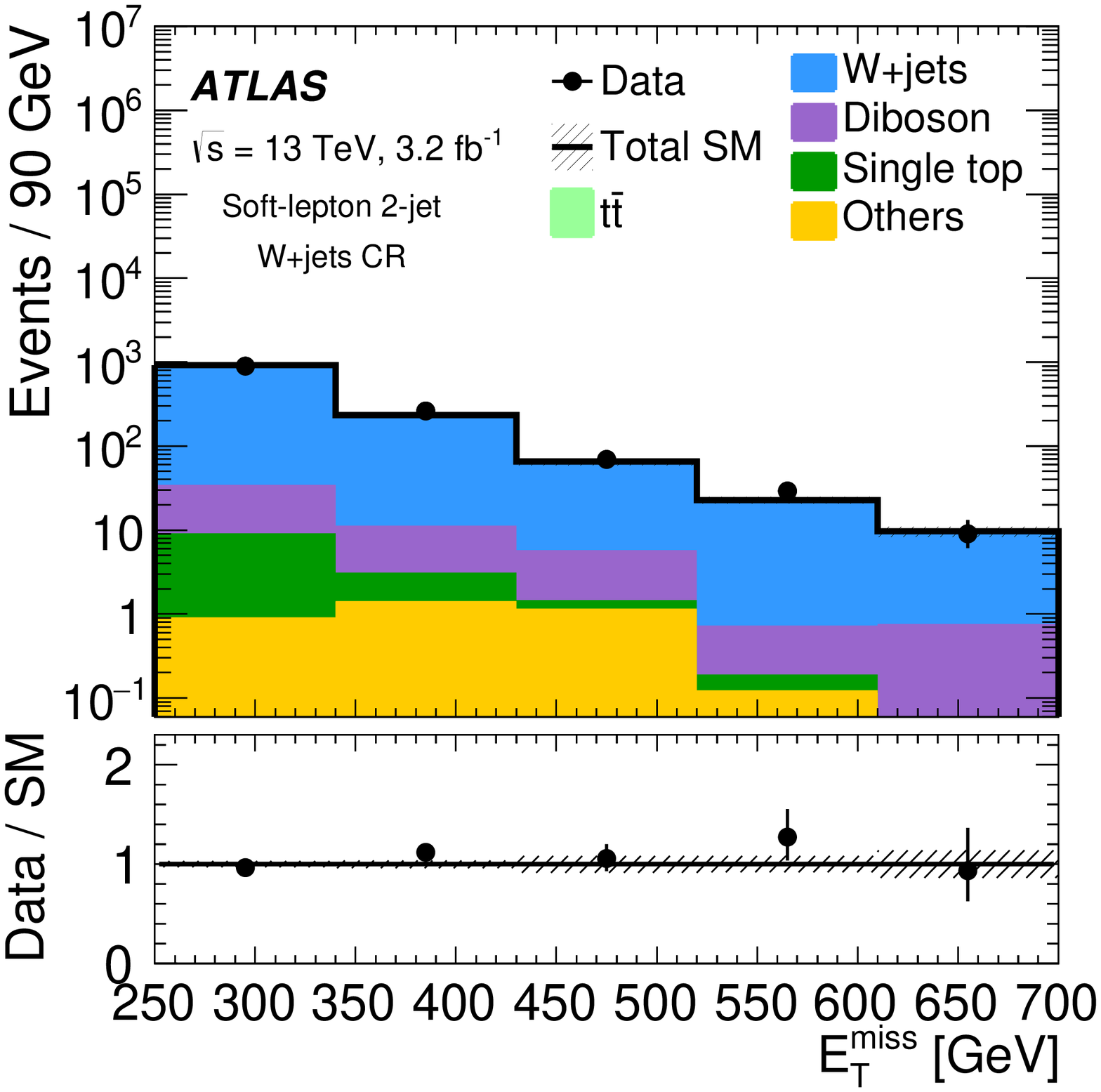}
  \end{minipage}
 \end{center}
\caption{\label{sec:background:CRplots}The distribution of the missing
transverse momentum is shown in hard-lepton 6-jet \ttbar\ (top left) and
$W$+jets (top right) and in the soft-lepton 2-jet \ttbar\ (bottom left) and
$W$+jets (bottom right) control regions after normalising the \ttbar\ and
$W$+jets background processes in the simultaneous fit. In the soft-lepton 2-jet
plots, the upper bound on $\met$ defining the control region is not applied.
The lower panels of the plots show the ratio of the observed data to the total SM background expected from simulations scaled to the number of events observed in the data. The uncertainty bands include all statistical and systematic uncertainties on simulation, as discussed in Section~\ref{sec:systematics}. The component `Others' is the sum of $Z$+jets and \ttbar+V.}
\end{figure}

To gain confidence in the extrapolation from control to signal regions using
simulated event samples, the results of the simultaneous fit are cross-checked
in validation regions which are disjoint with both the control and signal regions.
The validation regions are designed to be kinematically close to the signal
regions, as shown in Fig.~\ref{fig:hardSR}, while expecting only a small contamination from the signal in the models considered in this search. The
validation regions are not used to constrain parameters in the fit, but provide
a statistically independent cross-check of the extrapolation.

This analysis uses two validation regions per signal region. In the hard-lepton channel, one of the validation regions is used to test
the extrapolation to larger \mt\ values, while the other validation region tests
the extrapolation to larger aplanarity values or, in the case of the 4-jet high-$x$ selection, to larger values in \met/$m_{\mathrm{eff}}^{\mathrm{incl}}$.
In the soft-lepton channel, the validation regions are used to test the extrapolation to larger \met, \mt\ or \HT\ values.

%
\section{Systematic uncertainties}
\label{sec:systematics}
Two categories of systematic uncertainties have an impact on the results
presented here: uncertainties arising from experimental effects, and
uncertainties associated with theoretical predictions and modelling. Their effects
are evaluated for all signal samples and background processes. Since the
normalisation of the dominant background processes is extracted in dedicated
control regions, the systematic uncertainties only affect the extrapolation to
the signal regions in these cases. 

Among the dominant experimental systematic uncertainties are the jet energy scale (JES) and resolution
(JER) and the muon momentum resolution. The jet uncertainties are derived as a
function of $\pt$ and $\eta$ of the jet, as well as of the pile-up conditions
and the jet flavour composition of the selected jet sample. They are determined
using a combination of simulated samples and studies of data, such as measurements of
the jet balance in dijet, $Z$+jet and $\gamma$+jet
events~\cite{ATL-PHYS-PUB-2015-015}. 
The $\Jpsi\to\Pleptonplus\Pleptonminus$, $W^{\pm}\to\Plepton^{\pm}\nu$ and $Z\to\Pleptonplus\Pleptonminus$ decays in data and simulation are exploited to estimate the uncertainties in lepton reconstruction, identification, momentum/energy scale and resolution, and isolation criteria~\cite{ATL-PHYS-PUB-2015-037,ATL-PHYS-PUB-2015-041}. In particular, muon momentum resolution and scale calibrations are derived for simulation from a template fit that compares the invariant mass of $Z\to\mu\mu$ and $\Jpsi\to\mu\mu$ candidates in data and simulation. The corresponding uncertainties are computed from variations of several fit parameters, following the procedure described in Ref.~\cite{Aad:2014rra}.

The simulation is reweighted to match the distribution of the average number of proton-proton interactions per bunch crossing observed in data.  In the signal regions characterised by a higher jet multiplicity, the uncertainty arising from this reweighting also becomes relevant.

The systematic uncertainties related to the modelling of $\met$ in the
simulation are estimated by propagating the uncertainties on the energy and
momentum scale of each of the objects entering the calculation, as well as the uncertainties on
the soft term resolution and scale.

Different uncertainties in the theoretical modelling of the SM production
processes are considered, as described in the following.

For $t\bar{t}$, single-top and $W/Z$+jets samples, the uncertainties related to
the choice of QCD renormalisation and factorisation scales are assessed by
varying the corresponding generator parameters up and down by a factor of two around their nominal values. Uncertainties in the resummation scale and
the matching scale between matrix elements and parton shower are evaluated for the $W$+jets samples by varying up and down by a factor of two the corresponding parameters in \sherpa.

For $t\bar{t}$ and single-top production, specific samples with an increased and decreased
amount of initial- and final-state radiation are compared to the nominal
sample. The relative difference in the extrapolation factors ($t\bar{t}$) or expected
rates (single-top) is assigned as an uncertainty. Moreover, the uncertainty
associated with the parton shower modelling is assessed as the difference between
the predictions from {\sc powheg+Pythia} and {\sc powheg+Herwig++}\footnote{The
{\sc{Herwig++}} version 2.7.1 used.}~\cite{Bahr:2008pv}.

An uncertainty arising from the choice of parton level generator is estimated for
$t\bar{t}$, diboson and $W/Z$+jets processes. In the former case, the
predictions from {\sc powheg-box} are compared to {\sc aMc@NLO}\footnote{The {\sc aMc@NLO} version
2.1.1 is used.}~\cite{Frixione:2002ik}; for dibosons, \sherpa\ is compared to
{\sc powheg-box}; for $W/Z$+jets, \sherpa\ is compared to {\sc Madgraph}~\cite{Alwall:2011uj}.

An uncertainty of $5$\% in the inclusive $Z$+jets cross-section is
assumed~\cite{ATLAS-CONF-2015-039}. Uncertainties in the inclusive single-top
cross-sections are assigned as $3.7$\% ($s$-channel, top), $4.7$\%
($s$-channel, anti-top), $4$\% ($t$-channel, top), $5$\% ($t$-channel, anti-top) and
$5.3$\% ($Wt$-channel)~\cite{Kant:2014oha}. Samples using diagram subtraction and diagram removal schemes are compared for assessing the sensitivity to the treatment of interference effects between single-top and \ttbar\ production at NLO.

An overall systematic uncertainty of $6$\% in the inclusive cross-section is
assigned to the small contribution from $WW$, $WZ$, $ZZ$, $W\gamma$ and $Z\gamma$
processes, which are estimated entirely from simulation. The uncertainty accounts for missing higher-order corrections, for the uncertainty in the value of the strong coupling constant, and for the uncertainties on the PDF sets. The uncertainties associated with the resummation, factorisation and renormalisation scales are computed by varying the corresponding \sherpa parameters.

For the very small contributions of $t\bar{t}+W/Z/WW$, an uncertainty of $30$\% is assigned. 

Among the main systematic uncertainties on the total background predictions in
the various signal regions are the ones associated with the finite size of the MC samples, which range from 11\% in the hard-lepton 6-jet SR to 33\% in the hard-lepton 4-jet high-$x$ SR. Moreover, the uncertanties associated with the normalisation
of the \ttbar\ background, ranging from $7$\% in the soft-lepton 5-jet SR to $21$\%
in the hard-lepton 4-jet low-$x$ SR. Further important uncertainties are the theoretical uncertainties
associated with the single-top background in the hard-lepton regions, which amount to $3$\% in the 4-jet high-$x$ signal region and increase to as much as $34$\% in
the 4-jet low-$x$ signal region, and the theoretical uncertainties on the $W$+jets background in the soft-lepton regions
(up to $11$\%).

The theoretical systematic uncertainty affecting the modelling of ISR can become
sizeable in the simplified signal models used in this analysis, especially when the SUSY
particles' mass splitting becomes small. Variations of a factor of two in
the following {\sc Madgraph} and {\sc Pythia} parameters are used to
estimate these uncertainties: the renormalisation and factorisation scales, the initial- and the final-state radiation scales, as well as the {\sc Madgraph} jet matching scale. The overall uncertainties range from about $5$\% for signal models with large mass differences between the gluino, the chargino, and the neutralino, to $25$\% for models with very compressed mass spectra.

%
\section{Statistical analysis}
\label{sec:fit}

The final results are based on a profile likelihood method~\cite{Cowan:2010js}
using the HistFitter framework~\cite{Baak:2014wma}. To obtain a set of
background predictions that is independent of the observation in the signal
regions, the fit can be configured to use only the control regions to constrain
the fit parameters; this is referred to as the background-only fit. For each
signal region a background-only fit is performed, based on the following inputs:

\begin{itemize}
 \item the observed number of events in each of the control regions associated with the signal region, together with the number of events expected from simulation;
 \item the extrapolation factors, including uncertainties, from control regions to the signal region, as obtained from simulation, for the $W$+jets and the \ttbar\ backgrounds;
 \item the yields of the smaller backgrounds such as the single-top, \ttbar+$V$,
 $Z$+jets and diboson backgrounds as obtained from simulation, including uncertainties.
\end{itemize}

Using this information a likelihood is constructed for every background-only
fit. It consists of a product of Poisson probability density functions for every
region and of constraint terms for systematic uncertainties as described below.

Multiple parameters are included in each likelihood: two normalisation
parameters describing the normalisation of the $W$+jets and \ttbar\ backgrounds
and nuisance parameters associated with the systematic uncertainties (as described
in Section~\ref{sec:systematics}) or the statistical uncertainties in simulated
event yields. The nuisance parameters associated with the systematic uncertainties
are constrained by Gaussian functions with their widths corresponding to the
size of the uncertainty, while the statistical uncertainties are constrained by
Poisson functions.
The parameters are correlated between the control regions and the signal region.

In the fit, the likelihood is
maximised by adjusting normalisation and nuisance parameters. The normalisation
scale factor of the \ttbar\ background is fitted to values between
$0.34^{+0.28}_{-0.25}$ (4-jet high-$x$ control regions) and
$0.92^{+0.14}_{-0.12}$ (5-jet soft-lepton control regions), the normalisation of
the $W$+jets background to values between $0.72^{+0.31}_{-0.33}$ (6-jet control
regions) and $1.00 \pm 0.04$ (2-jet soft-lepton control regions). Previous
analyses~\cite{Aad:2015mia} also found normalisation factors considerably
smaller than unity for these background processes in similarly extreme regions of phase space.
The fit introduces correlations between the normalisation parameters
associated with the \ttbar\ and the $W$+jets backgrounds and the nuisance parameters associated with systematic uncertainties.
The uncertainty in the total background estimate may thus be smaller or larger than the
sum in quadrature of the individual uncertainties.

%
\section{Results}
\label{sec:result}

The results of the background-only fit described in Section~\ref{sec:fit} in the
validation and signal regions are shown in Fig.~\ref{results:SRpull} and are
further detailed for the signal regions in Table~\ref{results:tab:yieldsSRs}.
Good agreement between predicted and observed event yields is seen in all
validation regions.

\begin{figure}
\begin{center}
\includegraphics[width=0.98\textwidth]{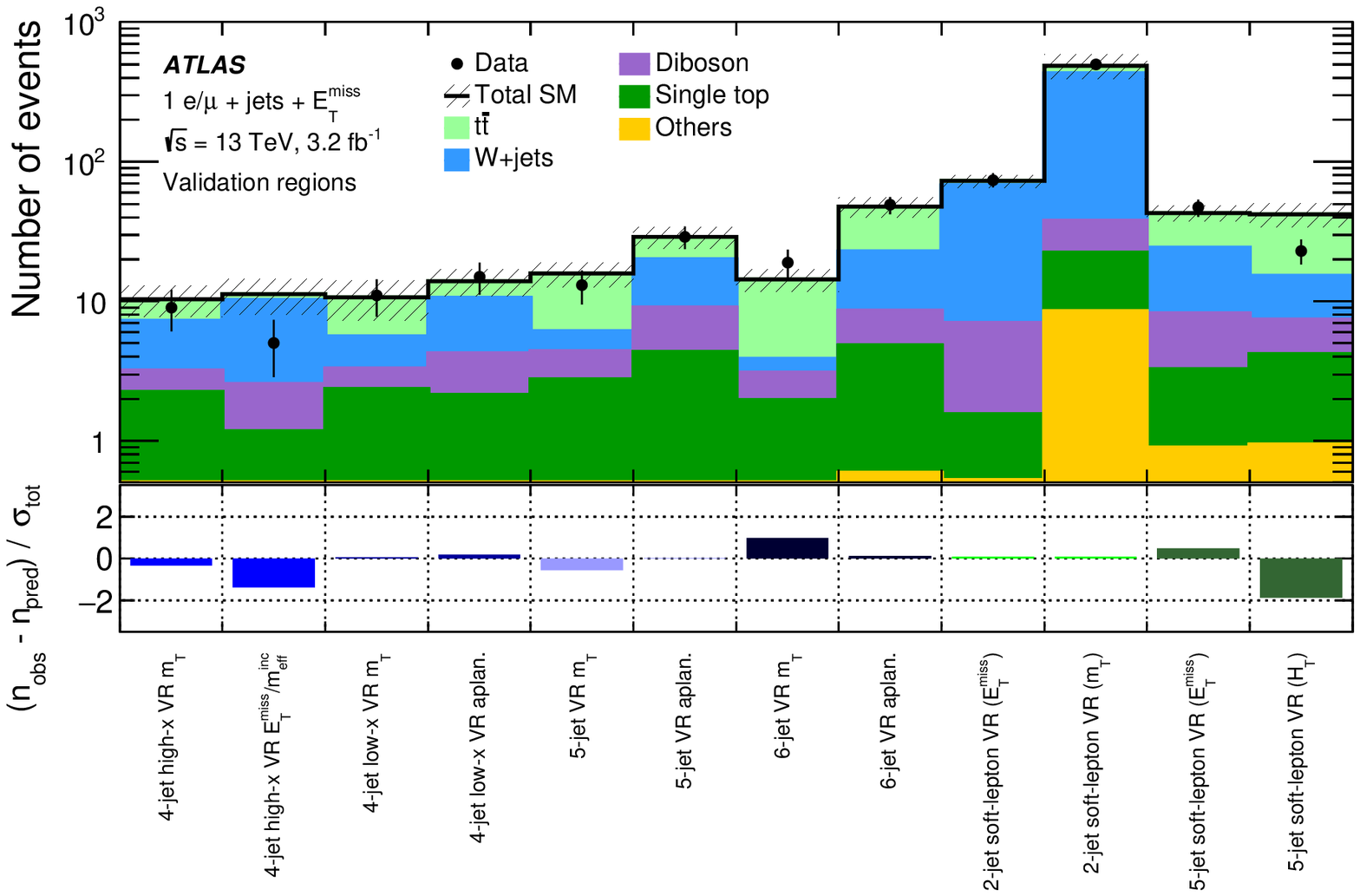}\\
  \includegraphics[width=0.98\textwidth]{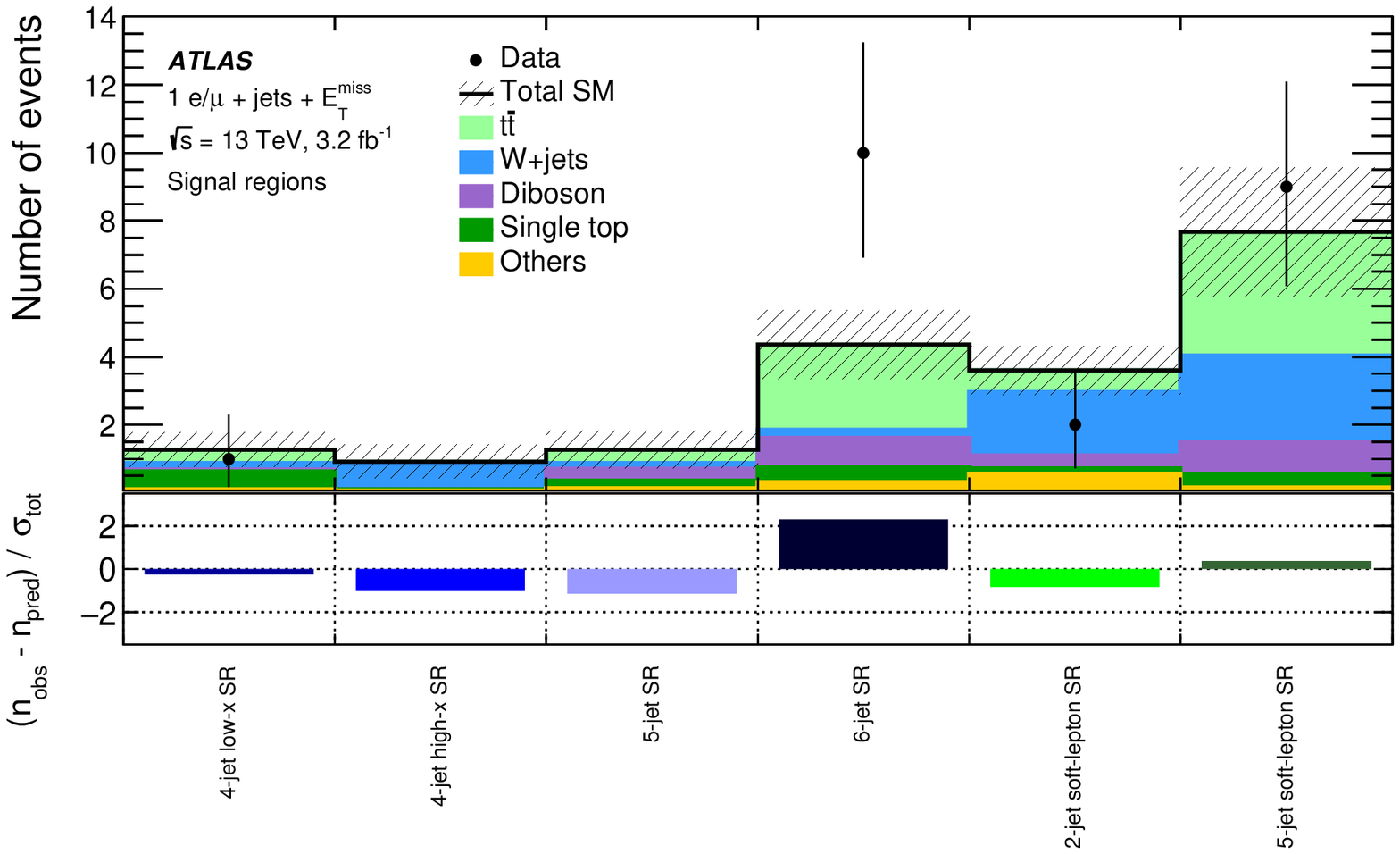}
\end{center}
\caption{\label{results:SRpull} Expected background yields as obtained in the background-only fits in all hard-lepton and soft-lepton validation (top plot) and signal (bottom plot) regions together with observed data are given in the top parts of the plots. Uncertainties in the fitted background estimates combine statistical (in the simulated event yields) and systematic uncertainties.
The bottom parts of the plots show the differences between observed ($n_\mathrm{obs}$) and predicted $n_\mathrm{pred}$ event yields, divided by the total (statistical and systematic) uncertainty in the prediction ($\sigma_\mathrm{tot}$). Bars sharing the same colour belong to regions fitted in the same background-only fit.}
\end{figure}

Figure~\ref{sec:results:dataMC} shows the \mt, \met\ and
\met/$m_{\mathrm{eff}}^{\mathrm{incl}}$ distributions before applying the
requirement on the plotted variable in the signal regions.

\begin{table}[t]

\renewcommand{\arraystretch}{1.2}
\caption{\label{results:tab:yieldsSRs} Background fit results for the hard-lepton and soft-lepton signal regions, for an integrated luminosity of $3.2$~\ifb.  
Uncertainties in the fitted background estimates combine statistical (in the simulated event yields) and systematic uncertainties. The uncertainties in this table are symmetrised for propagation purposes but truncated at zero to remain within the physical boundaries.
}
\begin{center}
{\footnotesize
\begin{tabular*}{\textwidth}{@{\extracolsep{\fill}}l|rrrr|rr}
\hline\hline
 & \multicolumn{4}{c|}{Hard-lepton} & \multicolumn{2}{c}{Soft-lepton}\\
 & 4-jet low $x$ & 4-jet high $x$ & 5-jet & 6-jet & 2-jet & 5-jet \\\hline\hline
 Observed events          & $1$   & $0$    & $0$  & $10$   & $2$ & $9$  \\
\hline
Fitted background events   & $1.3 \pm 0.5$    & $0.9 \pm 0.5$                & $1.3 \pm 0.6$     & $4.4 \pm 1.0$          & $3.6 \pm 0.7$     & $7.7 \pm 1.9$   \\
\hline
        \ttbar\    & $0.40 \pm 0.31$          & $0.08 \pm 0.07$              & $0.40 \pm 0.24$   & $2.5 \pm 0.9$          & $0.64 \pm 0.33$   & $3.6 \pm 1.2$ \\
        $W$+jets   & $0.19 \pm 0.12$          & $0.8 \pm 0.5$                & $0.16 \pm 0.12$   & $0.23 \pm 0.16$        & $1.9 \pm 0.5$     & $2.5 \pm 1.3$ \\
        $Z$+jets   & $0.045 \pm 0.023$        & $0.028 \pm 0.027$            & $0.073 \pm 0.035$   & $0.08 \pm 0.08$        & $0.47 \pm 0.12$   & $0.09 \pm 0.04$ \\
        Single-top & $0.5 \pm 0.5$            & $0.04^{+0.10}_{-0.04}$       & $0.21^{+0.22}_{-0.21}$  & $0.4 \pm 0.4$    & $0.16 \pm 0.14$   & $0.42 \pm 0.33$ \\
        Diboson    & $0.06^{+0.20}_{-0.06}$   & $0.002^{+0.014}_{-0.002}$    & $0.37 \pm 0.23$   & $0.9 \pm 0.5$          & $0.38 \pm 0.16$   & $0.9 \pm 0.6$ \\
        \ttbar+V   & $0.048 \pm 0.021$        & $0.024 \pm 0.012$            & $0.059 \pm 0.029$ & $0.23 \pm 0.08$        & $0.085 \pm 0.028$ & $0.065 \pm 0.024$ \\
\hline\hline
\end{tabular*}
}
\end{center}
\end{table}

\begin{figure}
 \begin{center}
  \begin{minipage}{0.43\textwidth}
   \includegraphics[width=\textwidth]{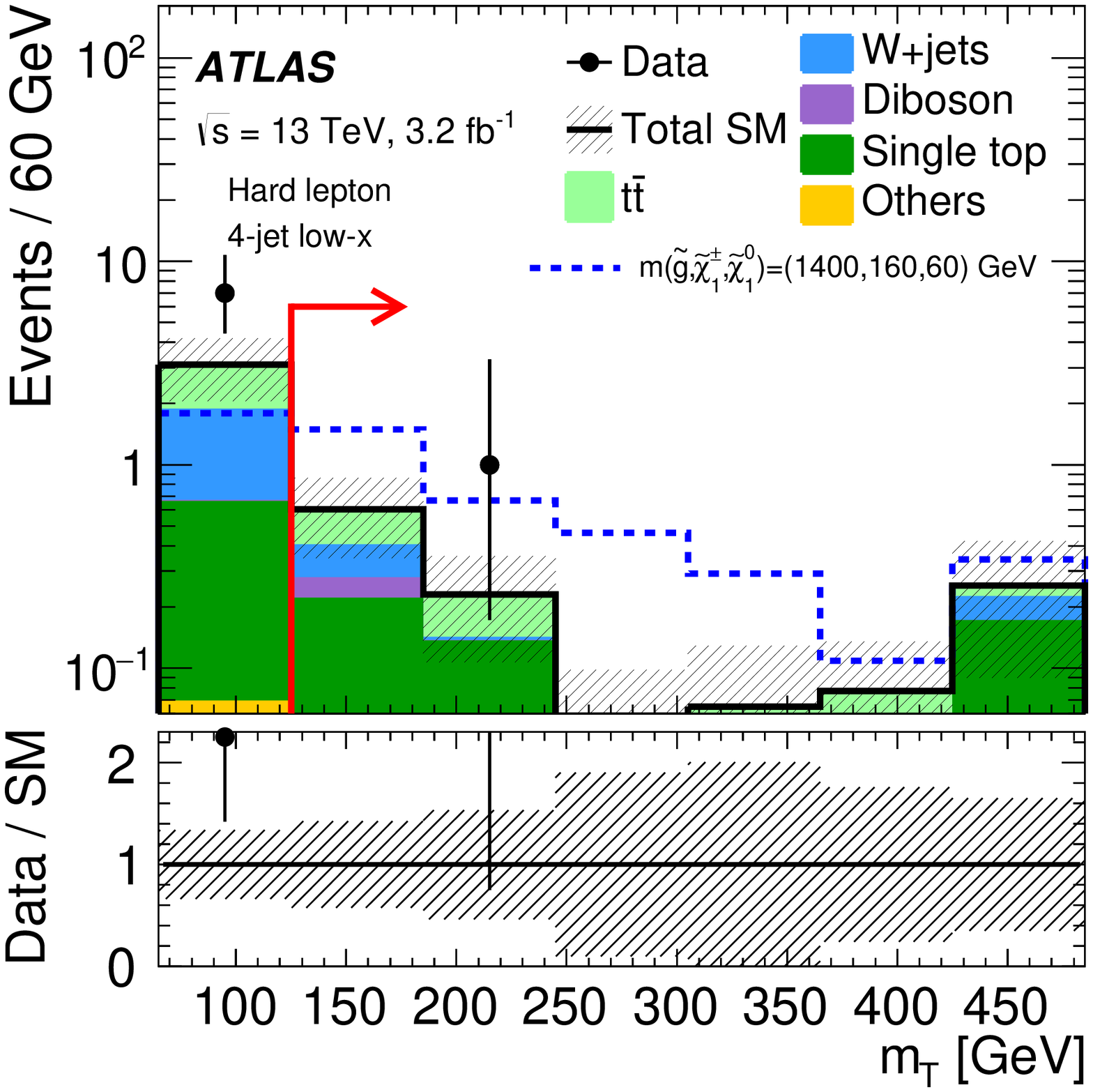}
  \end{minipage}
  \begin{minipage}{0.43\textwidth}
   \includegraphics[width=\textwidth]{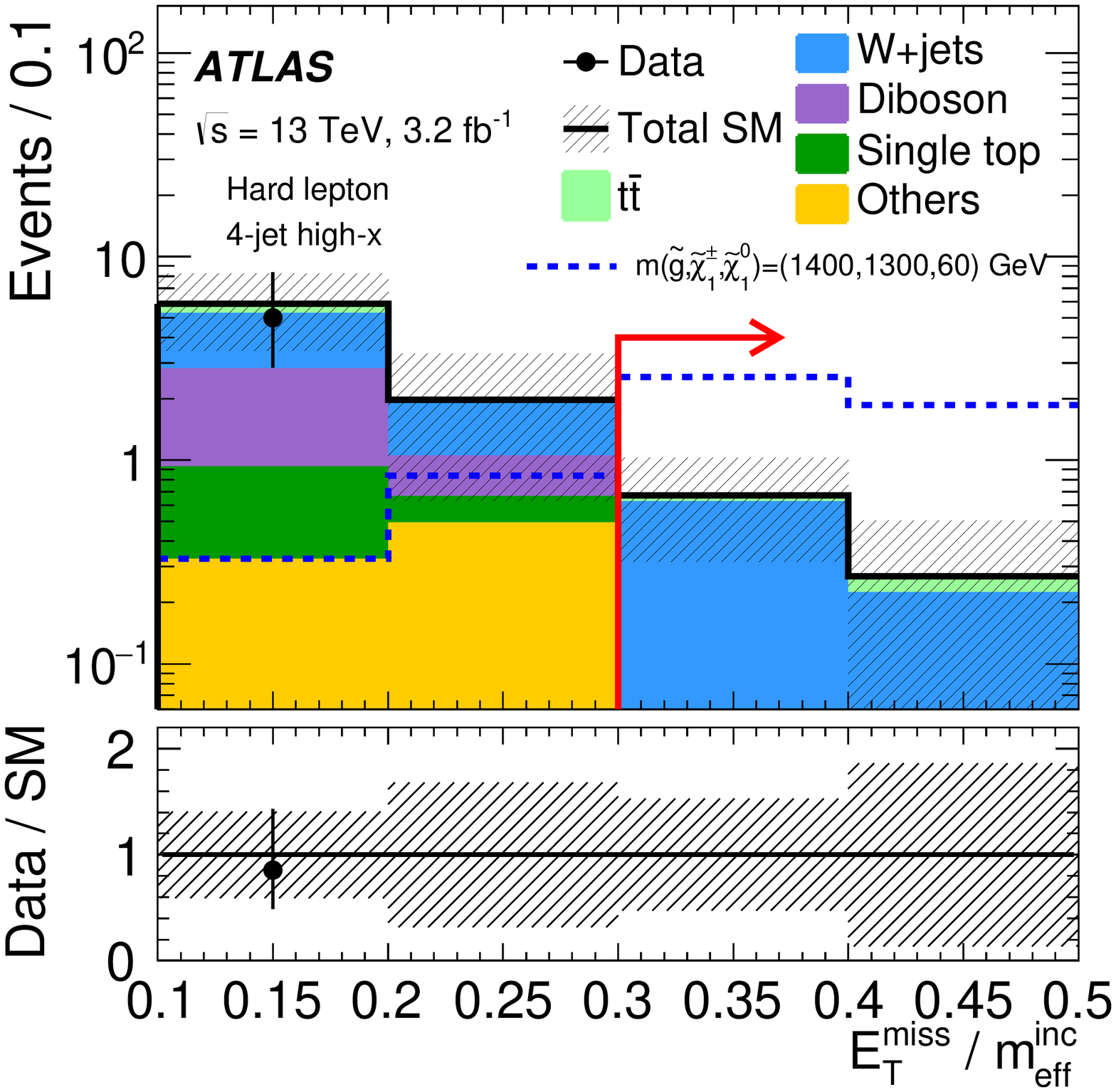}   
  \end{minipage}
    \begin{minipage}{0.43\textwidth}
   \includegraphics[width=\textwidth]{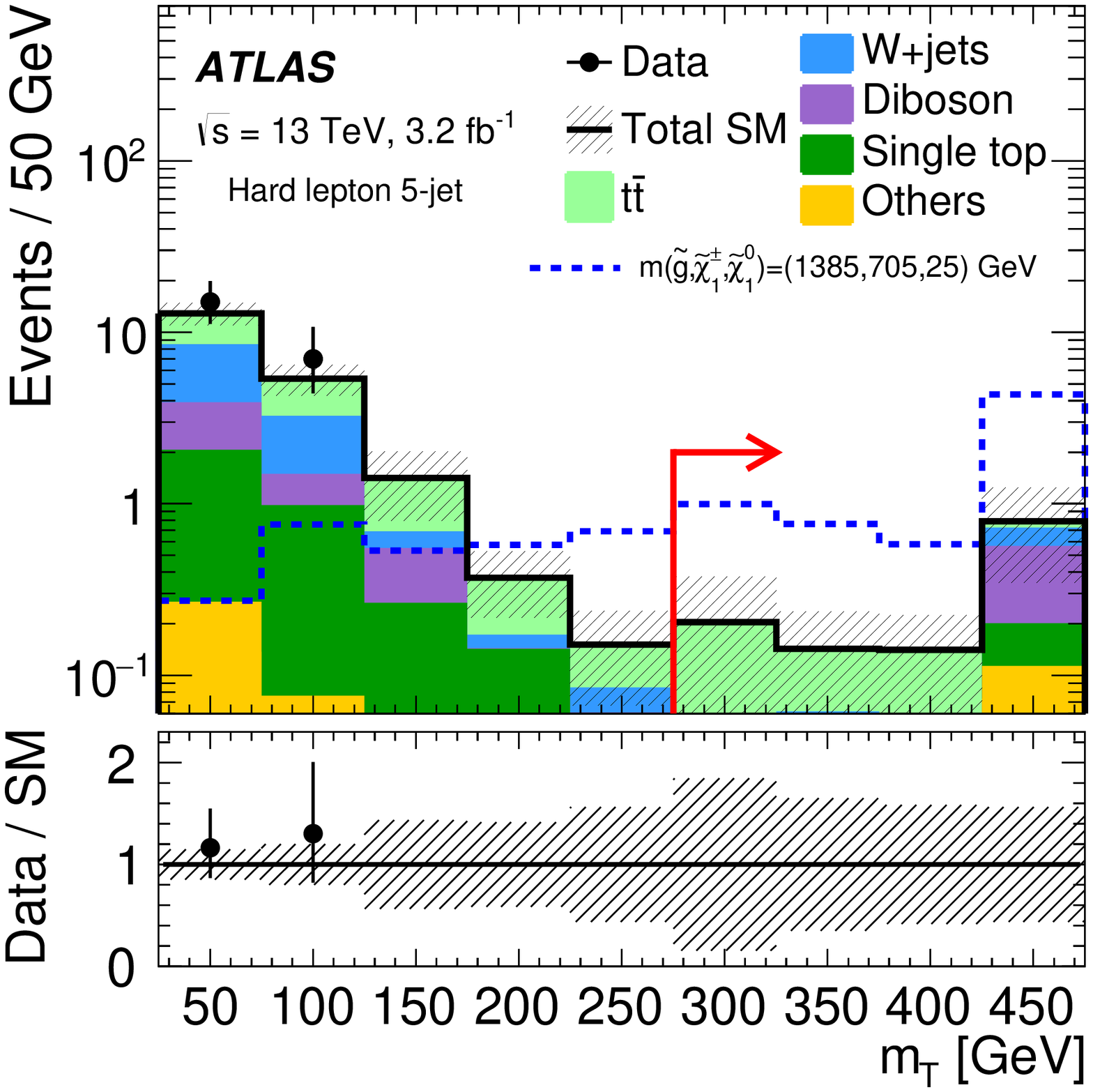}   
  \end{minipage}
    \begin{minipage}{0.43\textwidth}
   \includegraphics[width=\textwidth]{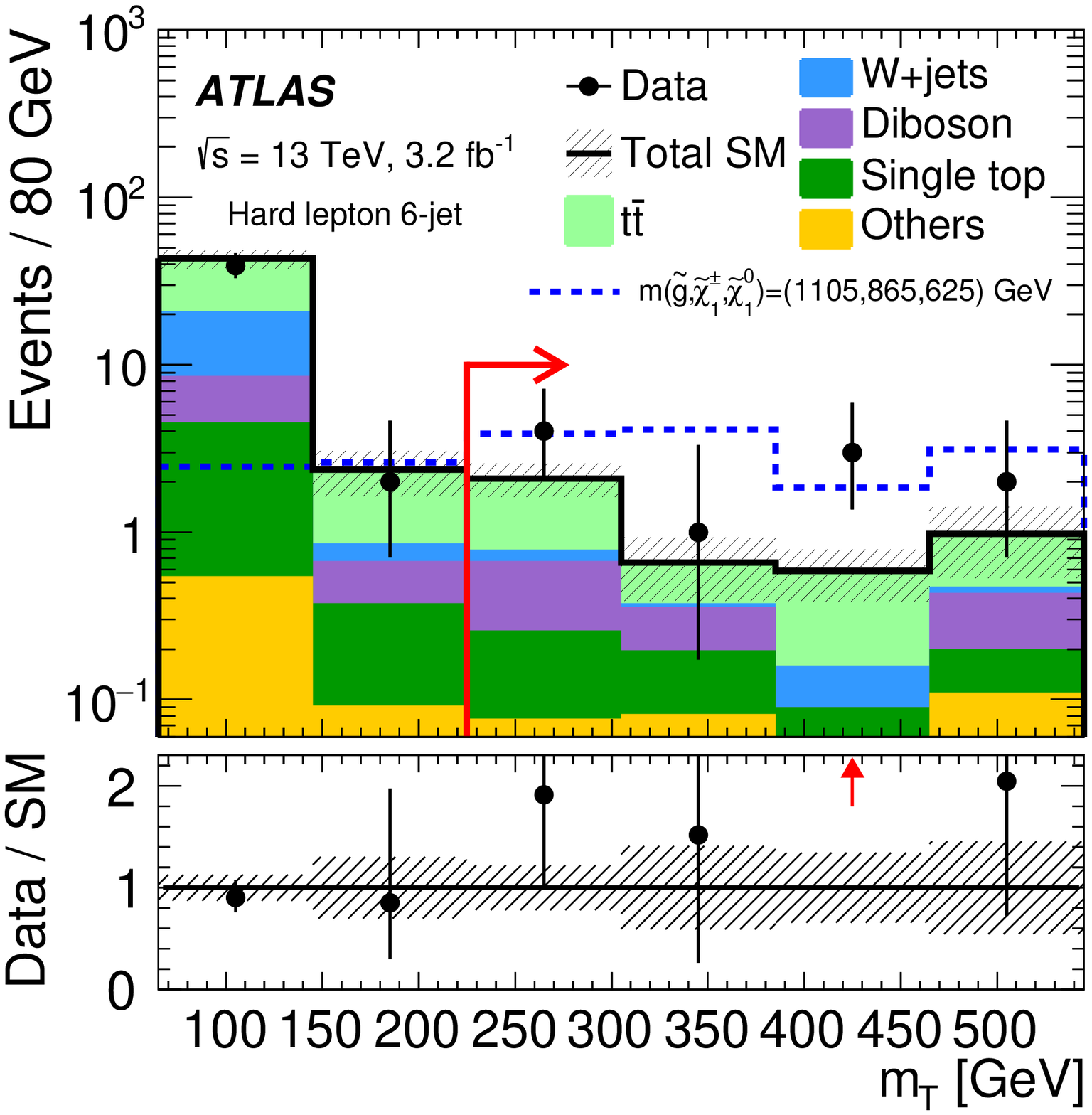}   
  \end{minipage}
    \begin{minipage}{0.43\textwidth}
   \includegraphics[width=\textwidth]{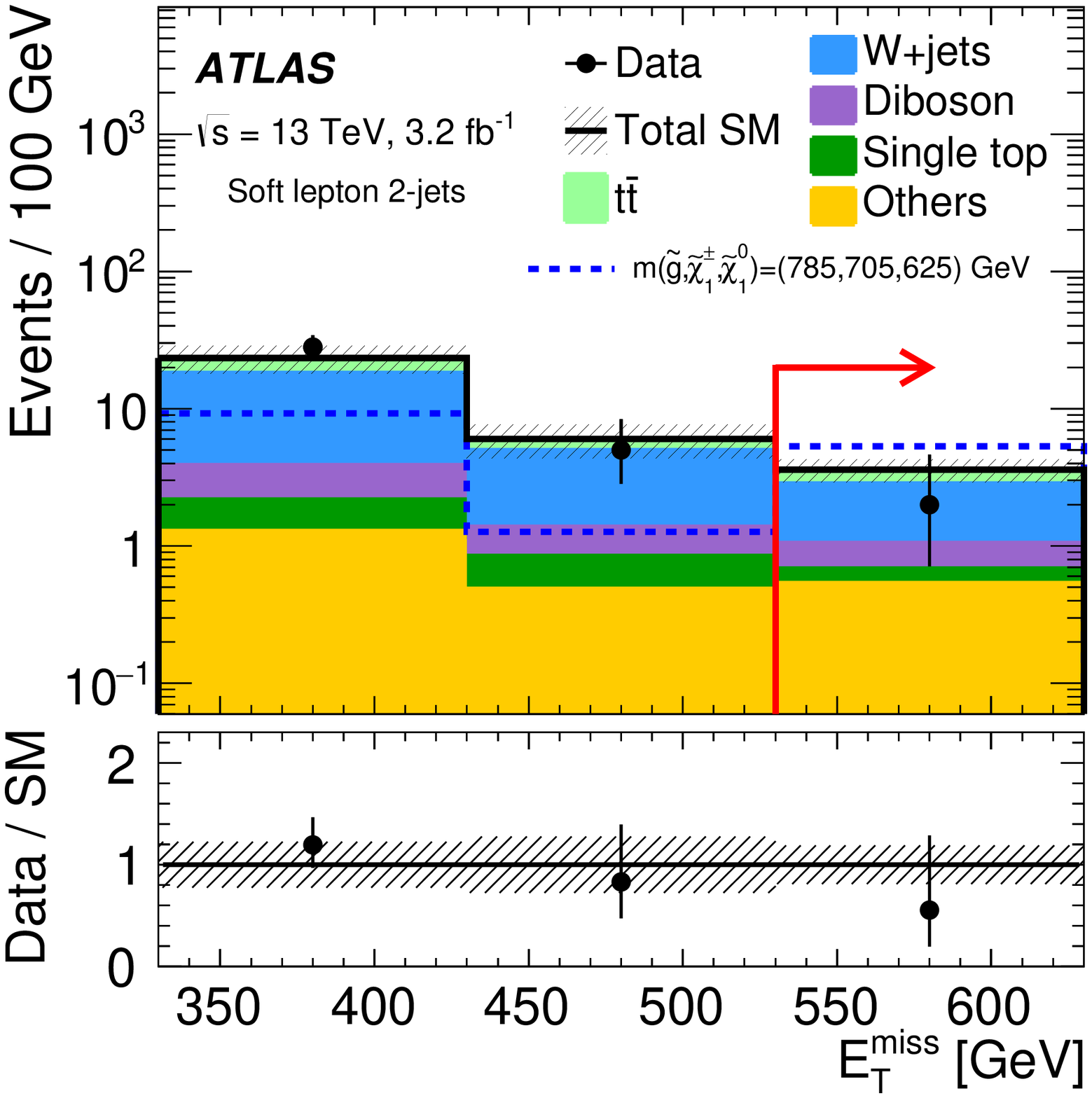}
  \end{minipage}
    \begin{minipage}{0.43\textwidth}
   \includegraphics[width=\textwidth]{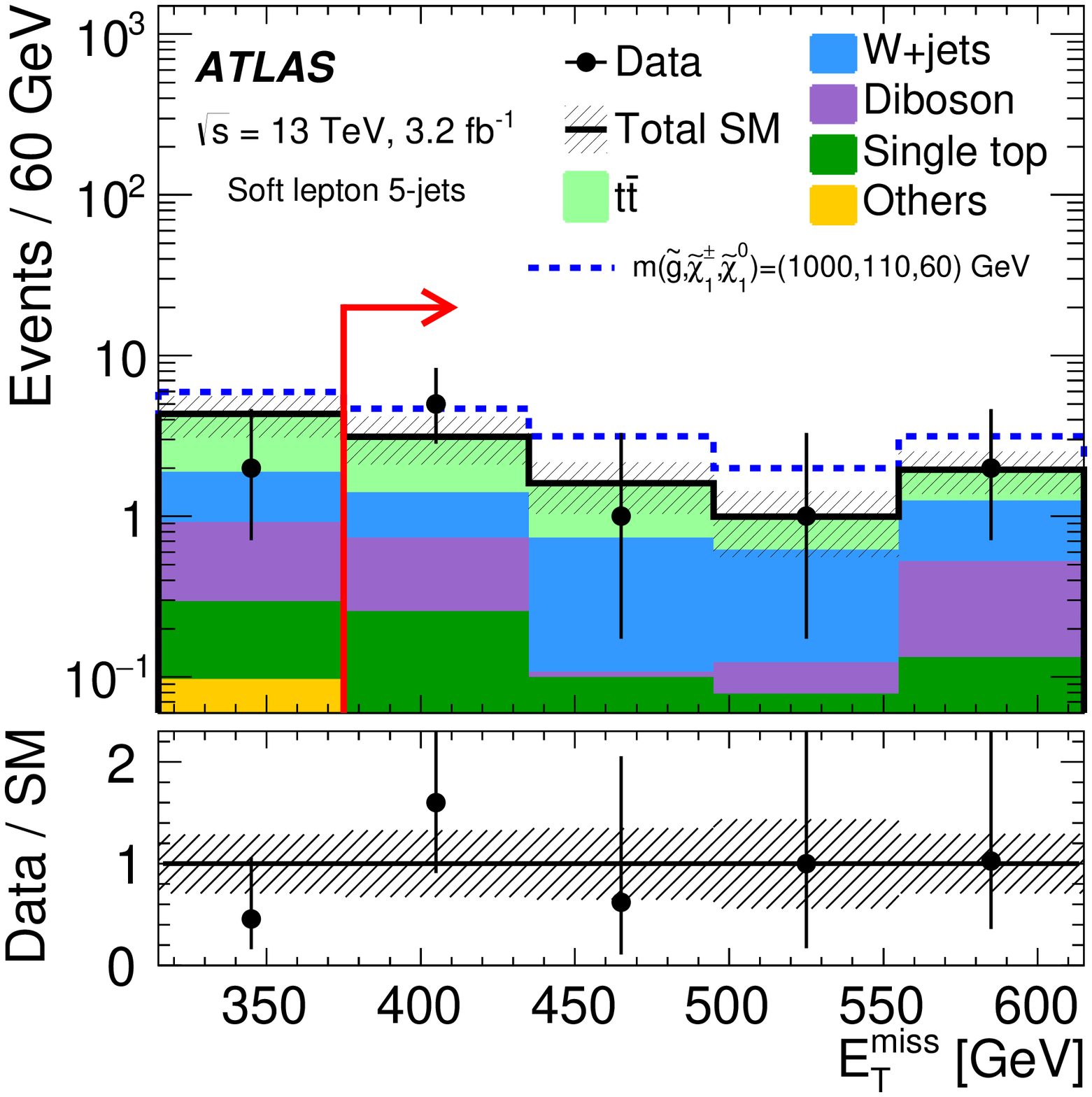}  
  \end{minipage}
 \end{center}
\caption{\label{sec:results:dataMC} Distributions of \mt\ for the hard-lepton 4-jet low-$x$ (top left), 5-jet (middle left), 6-jet (middle right) signal regions, of \met/$m_{\mathrm{eff}}^{\mathrm{incl}}$ for the 4-jet high-$x$ signal region (top right) and of \met\ for the soft-lepton 2-jet (bottom left) and soft-lepton 5-jet (bottom right) signal regions. The requirement on the variable plotted is removed from the definitions of the signal regions, where the arrow indicates the position of the cut in the signal region. The lower panels of the plots show the ratio of the observed data to the total background prediction as derived in the background-only fit. The uncertainty bands plotted include all statistical and systematic uncertainties as discussed in Section~\ref{sec:systematics}. The component `Others' is the sum of $Z$+jets and \ttbar+V. The last bin includes the overflow.
}
\end{figure}

The predicted background yields and the observed number of events agree in all
signal regions.
The largest deviation, $2.1$ standard deviations, is observed in the 6-jet hard-lepton signal region.
This excess arises only from the muon channel, in which $8$
events are observed, while $2.5 \pm 0.7$ events are predicted (local significance of
$2.6$ standard deviations). 
The electron channel shows good agreement, with $2$
events observed and $1.9 \pm 0.6$ predicted.

Model-independent upper limits and discovery $p$-values~\cite{Baak:2014wma} in the signal
regions are calculated in a modified fit configuration with respect to the background-only fit. The only region considered in these fits is the respective signal region. Control regions are not explicitly included and thus any signal contamination in the control
regions is not taken into account, thus giving conservative limits. These fits use the background estimates as derived in the
background-only fits as input and allow for a non-negative signal contribution in the signal region. An additional normalisation
parameter for the signal contribution is included.

Observed and expected upper limits at $95$\% confidence level (CL) on the number
of events signifying new phenomena beyond the SM ($S_\mathrm{obs}^{95}$ and
$S_\mathrm{exp}^{95}$, respectively) are derived based on the $CL_s$
prescription~\cite{Read:2002hq} and are shown in
Table~\ref{results:tab:upperlimits} together with the upper limits on the
visible beyond the SM cross-section ($\sigma_\mathrm{vis}$, defined as the product
of acceptance, selection efficiency and production cross-section). The latter is
calculated by dividing the observed upper limit on the beyond-SM events by the
integrated luminosity of $3.2$ \ifb. The table also gives the background-only
confidence level $CL_b$.

Table~\ref{results:tab:upperlimits} also shows the discovery $p$-values,
giving the probability for the background-only assumption to produce event
yields greater or equal to the observed data. The $CL_b$ and $p$-values use
different definitions of test statistics in their calculation, the former
with the signal-strength parameter set to one and the latter to zero.

\begin{table}
\caption[Breakdown of upper limits.]{
The columns show from left to right: the name of the respective signal region; the $95$\% confidence level (CL) upper limits on the visible cross-section
($\langle\epsilon\sigma\rangle_{\rm obs}^{95}$) and on the number of
signal events ($S_{\rm obs}^{95}$ ); 
the 95\% CL upper limit on the number of
signal events ($S_{\rm exp}^{95}$), given the expected number (and $\pm 1\sigma$
variations on the expectation) of background events;
the two-sided $CL_b$ value, i.e. the confidence level observed for
the background-only hypothesis and the one-sided discovery $p$-value ($p(s = 0)$). The discovery $p$-values are capped to $0.5$ in the case of observing less events than the fitted background estimates.
\label{results:tab:upperlimits}}
\begin{center}
\begin{tabular*}{\textwidth}{@{\extracolsep{\fill}}lccccc}
\noalign{\smallskip}\hline\noalign{\smallskip}
{\bf Signal region}                        & $\langle\epsilon{\rm \sigma}\rangle_{\rm obs}^{95}$[fb]  &  $S_{\rm obs}^{95}$  & $S_{\rm exp}^{95}$ & $CL_{b}$ & $p(s=0)$  \\
\noalign{\smallskip}\hline\noalign{\smallskip}
Hard-lepton & & & & & \\
\noalign{\smallskip}\hline\noalign{\smallskip}
4-jet low-$x$   & $1.23$ &  $3.9$ & $ { 4.1 }^{ +1.5 }_{ -0.9 }$ & $0.46$ & $ 0.50$ \\%
4-jet high-$x$   & $0.87$ &  $2.8$ & $ { 2.9 }^{ +1.3 }_{ -0.2 }$ & $0.27$ & $ 0.50$ \\%
5-jet  & $0.87$ &  $2.8$ & $ { 3.5 }^{ +1.4 }_{ -0.7 }$ & $0.19$ & $ 0.50$ \\%
6-jet   & $3.90$ &  $12.5$ & $ { 6.5 }^{ +2.6 }_{ -1.6 }$ & $0.98$ & $ 0.02$ \\%
\noalign{\smallskip}\hline\noalign{\smallskip}
Soft-lepton & & & & & \\
\noalign{\smallskip}\hline\noalign{\smallskip}
2-jet  & $1.33$ &  $4.3$ & $ { 5.3 }^{ +2.2 }_{ -1.3 }$ & $0.23$ & $ 0.50$ \\%
5-jet   & $2.87$ &  $9.2$ & $ { 8.1 }^{ +2.9 }_{ -2.1 }$ & $0.68$ & $ 0.34$ \\%
\noalign{\smallskip}\hline\noalign{\smallskip}
\end{tabular*}
\end{center}
\end{table}

Model-dependent limits are calculated in a modified fit configuration with
respect to the background-only fit. A signal contribution is allowed and
considered in all control and signal regions, with a non-negative signal-strength normalisation parameter included. For the signal processes, uncertainties due to detector effects and theoretical modelling are considered. The signal regions are explicitly used in the fit to constrain the likelihood parameters.
Figure~\ref{sec:results:contours} shows the combined $95$\% CL exclusion limits in the simplified
models with gluino production using for each model point the signal region with the best expected
sensitivity. Gluino masses up to $1.6$ \TeV\
are excluded for scenarios with large mass differences between the gluino and
the neutralino and $x = (m_{\tilde{\chi}_{1}^{\pm}} - m_{\tilde{\chi}^{0}_{1}})/(m_{\tilde{g}} -
m_{\tilde{\chi}^{0}_{1}}) = 1/2$. In the same scenario and for models with a small mass
difference between the gluino and the neutralino, gluino masses up to $870$ \GeV\ are excluded. 
The signal regions address very different sets of models and are complementary to each other. In the case of the hard-lepton 6-jet signal region (covering the central part in the ($m_{\tilde{g}},m_{\tilde{\chi}^{0}_{1}}$) mass plane in Fig.~\ref{sec:results:contours}), the observed exclusion limit is considerably weaker than the expected one due to the excess seen in this region. 

\begin{figure}
 \begin{center}
  \includegraphics[width=14cm]{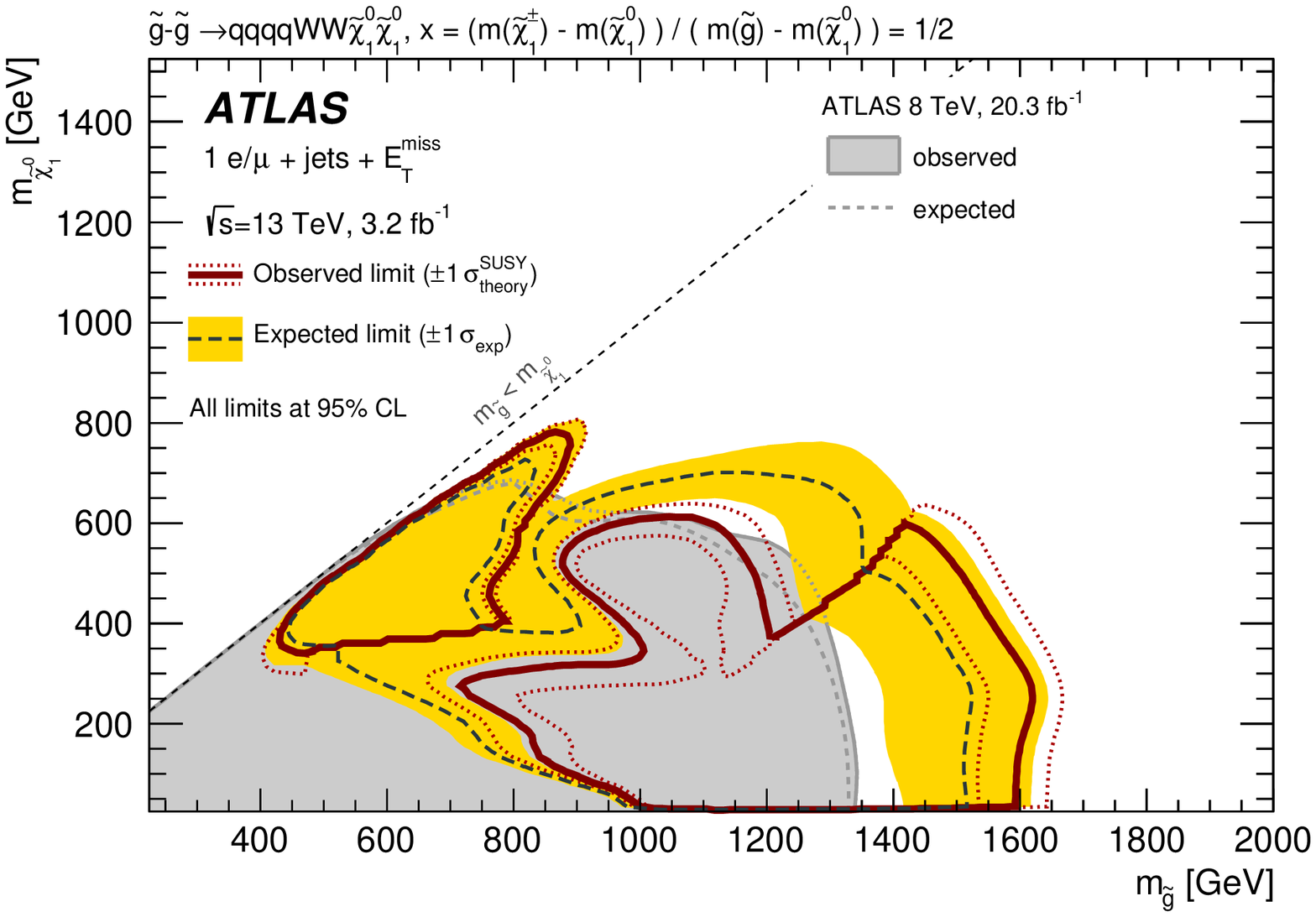}
  \includegraphics[width=14cm]{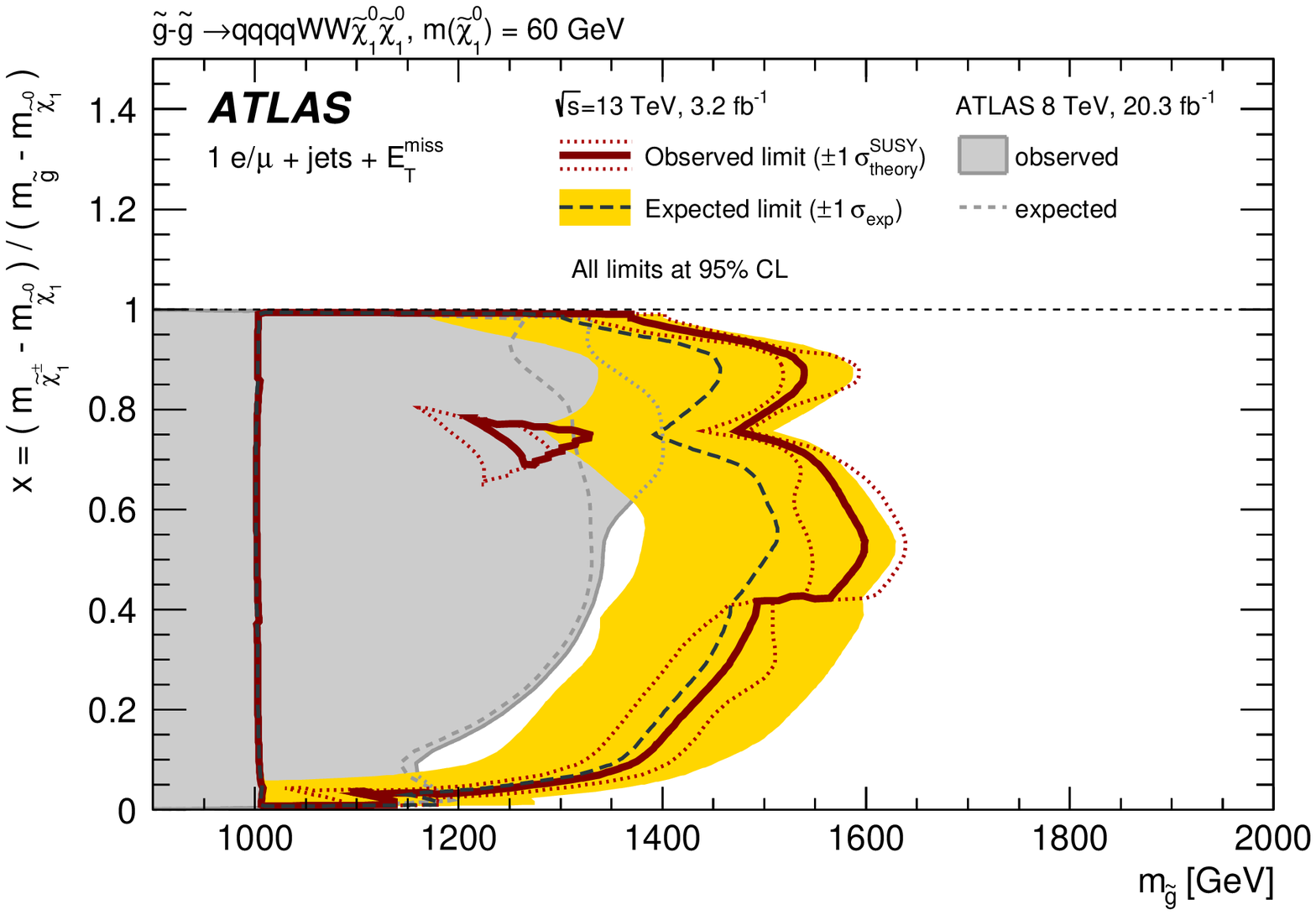}
 \end{center}
\caption{
\label{sec:results:contours}
Combined $95$\% CL exclusion limits in the two gluino simplified
models using for each model point the signal region with the best expected
sensitivity. The limits are presented in the
($m_{\tilde{g}},m_{\tilde{\chi}^{0}_{1}}$) mass plane (top) for the
scenario where the mass of the chargino $\tilde{\chi}_{1}^{\pm}$ is fixed to $x
= (m_{\tilde{\chi}_{1}^{\pm}} - m_{\tilde{\chi}^{0}_{1}})/(m_{\tilde{g}} -
m_{\tilde{\chi}^{0}_{1}}) = 1/2$ and in the ($m_{\tilde{g}},x$) plane
(bottom) for the $m_{\tilde{\chi}^{0}_{1}} = 60 \GeV$ models. The red solid line
corresponds to the observed limit with the red dotted lines indicating the
$\pm 1 \sigma$ variation of this limit due to the effect of theoretical scale and PDF
uncertainties in the signal cross-section. The dark grey dashed line indicates
the expected limit with the yellow band representing the $\pm 1 \sigma$ variation of the median expected limit due to the experimental and theoretical uncertainties. The exclusion limits at $95$\% CL by
previous ATLAS analyses~\cite{Aad:2015iea} are shown as the grey area.}
\end{figure}

\section{Conclusion}
\label{sec:conclusion}

A search for gluinos in events with one isolated lepton, jets, and missing
transverse momentum is presented. The analysis uses $3.2~\ifb$ of
proton--proton collision data collected by the ATLAS experiment in 2015 at
$\sqrt{s} = 13 \TeV$ at the LHC. Six signal regions requiring at least
two to six jets are used
to cover a broad spectrum of the targeted SUSY model parameter space. While four
signal regions are based on high-$\pT$ lepton selections and target models with large mass differences 
between the supersymmetric particles, two dedicated low-$\pT$ lepton regions are
designed to enhance the sensitivity to models with compressed mass spectra.

The observed data agree with the Standard Model background
prediction in the signal regions. The largest deviation is a $2.1$ standard
deviation excess in a channel requiring a high-$\pT$ lepton and six jets. For all signal regions, limits on the visible cross-section are 
derived in models of
new physics within the kinematic requirements of this search. In addition,
exclusion limits are placed on models with gluino production and subsequent
decays via an intermediate chargino to the lightest neutralino. The exclusion
limits of previous searches conducted in LHC Run 1 are significantly extended. Gluino masses up to $1.6~\TeV$ are excluded for low neutralino masses ($\lesssim 300 \GeV$) and chargino masses of $\sim 850 \GeV$.
\FloatBarrier

\section*{Acknowledgements}


We thank CERN for the very successful operation of the LHC, as well as the
support staff from our institutions without whom ATLAS could not be
operated efficiently.

We acknowledge the support of ANPCyT, Argentina; YerPhI, Armenia; ARC, Australia; BMWFW and FWF, Austria; ANAS, Azerbaijan; SSTC, Belarus; CNPq and FAPESP, Brazil; NSERC, NRC and CFI, Canada; CERN; CONICYT, Chile; CAS, MOST and NSFC, China; COLCIENCIAS, Colombia; MSMT CR, MPO CR and VSC CR, Czech Republic; DNRF and DNSRC, Denmark; IN2P3-CNRS, CEA-DSM/IRFU, France; GNSF, Georgia; BMBF, HGF, and MPG, Germany; GSRT, Greece; RGC, Hong Kong SAR, China; ISF, I-CORE and Benoziyo Center, Israel; INFN, Italy; MEXT and JSPS, Japan; CNRST, Morocco; FOM and NWO, Netherlands; RCN, Norway; MNiSW and NCN, Poland; FCT, Portugal; MNE/IFA, Romania; MES of Russia and NRC KI, Russian Federation; JINR; MESTD, Serbia; MSSR, Slovakia; ARRS and MIZ\v{S}, Slovenia; DST/NRF, South Africa; MINECO, Spain; SRC and Wallenberg Foundation, Sweden; SERI, SNSF and Cantons of Bern and Geneva, Switzerland; MOST, Taiwan; TAEK, Turkey; STFC, United Kingdom; DOE and NSF, United States of America. In addition, individual groups and members have received support from BCKDF, the Canada Council, CANARIE, CRC, Compute Canada, FQRNT, and the Ontario Innovation Trust, Canada; EPLANET, ERC, FP7, Horizon 2020 and Marie Sk{\l}odowska-Curie Actions, European Union; Investissements d'Avenir Labex and Idex, ANR, R{\'e}gion Auvergne and Fondation Partager le Savoir, France; DFG and AvH Foundation, Germany; Herakleitos, Thales and Aristeia programmes co-financed by EU-ESF and the Greek NSRF; BSF, GIF and Minerva, Israel; BRF, Norway; Generalitat de Catalunya, Generalitat Valenciana, Spain; the Royal Society and Leverhulme Trust, United Kingdom.

The crucial computing support from all WLCG partners is acknowledged gratefully, in particular from CERN, the ATLAS Tier-1 facilities at TRIUMF (Canada), NDGF (Denmark, Norway, Sweden), CC-IN2P3 (France), KIT/GridKA (Germany), INFN-CNAF (Italy), NL-T1 (Netherlands), PIC (Spain), ASGC (Taiwan), RAL (UK) and BNL (USA), the Tier-2 facilities worldwide and large non-WLCG resource providers. Major contributors of computing resources are listed in Ref.~\cite{ATL-GEN-PUB-2016-002}.

\printbibliography

\newpage 

\begin{flushleft}
{\Large The ATLAS Collaboration}

\bigskip

G.~Aad$^\textrm{\scriptsize 87}$,
B.~Abbott$^\textrm{\scriptsize 114}$,
J.~Abdallah$^\textrm{\scriptsize 65}$,
O.~Abdinov$^\textrm{\scriptsize 12}$,
B.~Abeloos$^\textrm{\scriptsize 118}$,
R.~Aben$^\textrm{\scriptsize 108}$,
M.~Abolins$^\textrm{\scriptsize 92}$,
O.S.~AbouZeid$^\textrm{\scriptsize 138}$,
H.~Abramowicz$^\textrm{\scriptsize 154}$,
H.~Abreu$^\textrm{\scriptsize 153}$
% ...
\bigskip
\\
% ...
$^{12}$ Institute of Physics, Azerbaijan Academy of Sciences, Baku, Azerbaijan\\
% ...
$^{65}$ University of Iowa, Iowa City IA, United States of America\\
% ...
$^{87}$ CPPM, Aix-Marseille Universit{\'e} and CNRS/IN2P3, Marseille, France\\
% ...
$^{92}$ Department of Physics and Astronomy, Michigan State University, East Lansing MI, United States of America\\
% ...
$^{108}$ Nikhef National Institute for Subatomic Physics and University of Amsterdam, Amsterdam, Netherlands\\
% ...
$^{114}$ Homer L. Dodge Department of Physics and Astronomy, University of Oklahoma, Norman OK, United States of America\\
% ...
$^{118}$ LAL, Univ. Paris-Sud, CNRS/IN2P3, Universit{\'e} Paris-Saclay, Orsay, France\\
% ...
$^{138}$ Santa Cruz Institute for Particle Physics, University of California Santa Cruz, Santa Cruz CA, United States of America\\
% ...
$^{153}$ Department of Physics, Technion: Israel Institute of Technology, Haifa, Israel\\
$^{154}$ Raymond and Beverly Sackler School of Physics and Astronomy, Tel Aviv University, Tel Aviv, Israel\\
% ...
\end{flushleft}


\end{document}